\documentclass[aps,prl,showpacs,superscriptaddress,twocolumn,amsmath,letter,amssymb,footinbib]{revtex4-1}

\usepackage[dvipdfmx]{graphicx, hyperref}
\hypersetup{colorlinks=true, linkcolor=red, citecolor = blue}
\usepackage{amsmath}

\usepackage{color}
\usepackage{cancel}
\usepackage{ulem}
\usepackage{comment}

\begin{document}
\title{Electro-mechano-optical detection of nuclear magnetic resonance}

\author{K.~Takeda}
\email{takezo@kuchem.kyoto-u.ac.jp}
\affiliation{Division of Chemistry, Graduate School of Science, Kyoto University, Kyoto 606-8502, Japan}
\author{K.~Nagasaka}
\author{A.~Noguchi}
\author{R.~Yamazaki}
\affiliation{Research Center for Advanced Science and Technology (RCAST), The University of Tokyo, Meguro-ku, Tokyo 153-8904, Japan}
\author{Y.~Nakamura}
\affiliation{Research Center for Advanced Science and Technology (RCAST), The University of Tokyo, Meguro-ku, Tokyo 153-8904, Japan}
\affiliation{Center for Emergent Matter Science (CEMS), RIKEN, Wako, Saitama 351-0198, Japan}
\author{E.~Iwase}
\affiliation{
Department of Applied Mechanics and Aerospace Engineering, Graduate School of Fundamental Science and Engineering, Waseda University, Shinjuku-ku, Tokyo 169-8555, Japan
}
\author{J.~M.~Taylor}
\affiliation{Research Center for Advanced Science and Technology (RCAST), The University of Tokyo, Meguro-ku, Tokyo 153-8904, Japan}
\affiliation{Joint Quantum Institute/NIST, College Park, Maryland 20742, USA}
\affiliation{Joint Center for Quantum Information and Computer Science, University of Maryland, College Park, Maryland 20742, USA}
\author{K.~Usami}
\email{usami@qc.rcast.u-tokyo.ac.jp}
\affiliation{Research Center for Advanced Science and Technology (RCAST), The University of Tokyo, Meguro-ku, Tokyo 153-8904, Japan}

\date{\today}

\begin{abstract}
Signal reception of nuclear magnetic resonance (NMR) usually relies on electrical amplification of the electromotive force caused by nuclear induction. Here, we report up-conversion of a radio-frequency NMR signal to an optical regime using a high-stress silicon nitride membrane that interfaces the electrical detection circuit and an optical cavity through the electro-mechanical and the opto-mechanical couplings. This enables optical NMR detection without sacrificing the versatility of the traditional nuclear induction approach. While the signal-to-noise ratio is currently limited by the Brownian motion of the membrane as well as additional technical noise, we find it can exceed that of the conventional electrical schemes by increasing the electro-mechanical coupling strength. The electro-mechano-optical NMR detection presented here opens the possibility of mechanical parametric amplification of NMR signals. Moreover, it can potentially be combined with the laser cooling technique applied to nuclear spins.

\begin{description}
\item[PACS numbers]12.20.Fv,42.50.Wk,42.50.Nn,76.60.-k
\end{description}
\end{abstract}

\maketitle

\section{Introduction}
\begin{figure*} [t]
\begin{center}
\includegraphics[width=\linewidth]{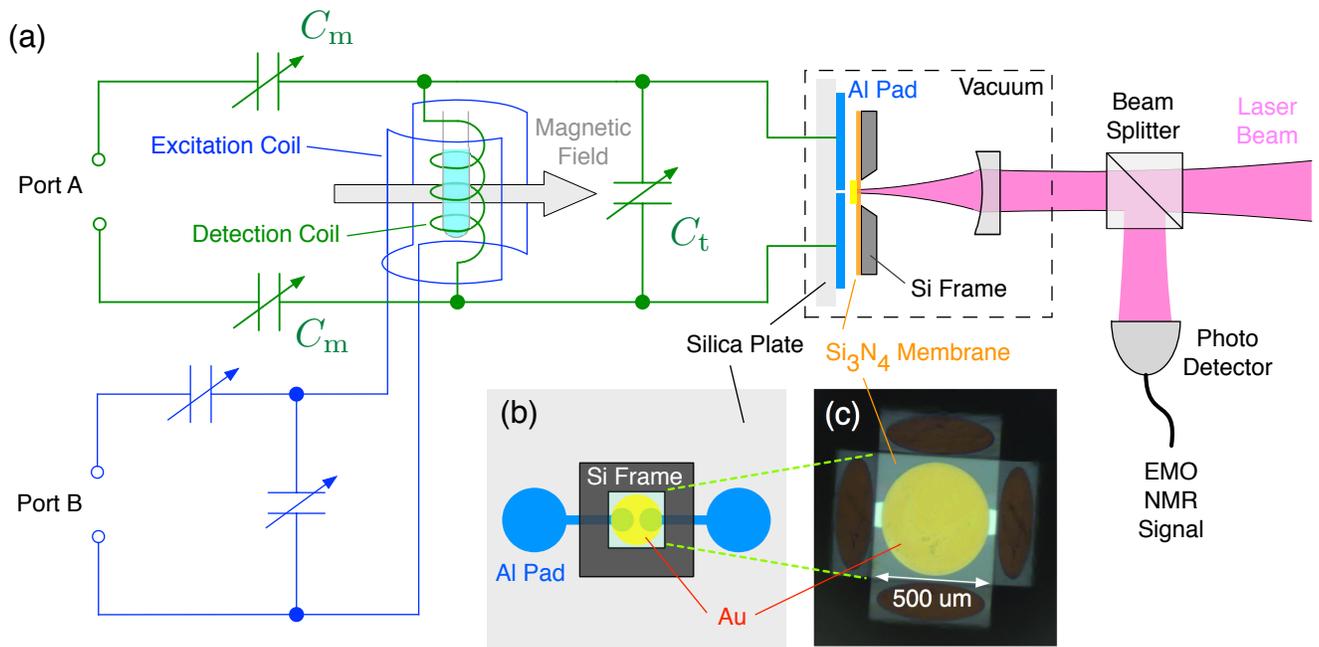}
\caption{(a) Experimental setup for EMO NMR composed of an orthogonal pair of coils tuned at the NMR frequency, a membrane put inside a vacuum chamber, an optical cavity, and a photo-detector. (b) Schematic drawing of the membrane capacitor. The Au layer on the membrane is electrically floating, and coupled capacitively to the Al pattern on the substrate. The two electrodes of the capacitor were electrically connected with the rest of the circuit through a pair of contact probes pushing against the Al pads on the silica substrate. (c) Photograph of the Au-deposited membrane.} \label{fig:setup}
\end{center}
\end{figure*}
Electrical signals can be up-converted from radio-frequency (rf) to optical regimes using a high-Q metal-coated silicon nitride membrane, which serves both as a capacitor electrode and a mirror of an optical interferometer~\cite{Polzik2014}.
There, the mechanics of the membrane, the electronics of the rf circuit, and the optics of the interferometer interact with one another through the opto-mechanical and the electro-mechanical couplings.
Even though the principle of such membrane-based, rf-to-light signal transduction has now been established, its power has yet to be harnessed in various rf-relevant fields.
In this work, we report on the first rf-to-light up-conversion of nuclear magnetic resonance (NMR) signals.

NMR~\cite{Bloch1946,Purcell1946,Purcell1948} is a powerful analytical tool, offering access to structure and dynamics in liquid and solid materials of physical/chemical/biological interest. Usually, NMR signal reception relies on nuclear induction~\cite{Bloch1946b} causing an electromotive force across the detection coil, followed by electrical amplification of the rf signals~\cite{HB1997}.
For a given signal strength, which could be significantly enhanced by nuclear hyperpolarization techniques~\cite{Abragam,Slichter2014}, the sensitivity is limited by the noises, namely, the Johnson noise of the resistive components within the circuit as well as the inevitable noise from the amplifier.
While the noise levels in unconventional optical NMR schemes, such as Faraday rotation~\cite{Kikkawa2000,Savukov2006}, force detection~\cite{PD2010}, fluorescence~\cite{Rugar2013,Wrachtrup2013}, and atomic magnetometry~\cite{Savukov2005}, are much lower than that in the traditional NMR and can in principle be quantum-noise-limited, all existing optical NMR detection schemes lack wide applicability compared to the traditional induction approach which allows measurements of any bulk samples, including living organisms, placed inside the detection coil.

Here, we put forward a \textit{versatile} approach to optical NMR readout, applicable straightforwardly to chemical analysis as well as magnetic reosnance imaging (MRI) diagnosis, by exploiting the membrane signal transducer system that we designed and fabricated to meet the specific needs for pulsed NMR spectroscopy.
In the following, we demonstrate the \textit{Electro-Mechano-Optical} (EMO) NMR detection scheme with proton ($^{1}$H) spin echoes~\cite{Hahn1950} in water. The signal-to-noise ratio, albeit currently limited by the thermal noise due to the Brownian motion of the membrane as well as additional technical noise, is expected to increase with the electro-mechanical coupling strength. We show that the EMO NMR approach can offer better sensitivity compared to the conventional all-electrical scheme with realistic improvements in the experimental parameters. The EMO approach opens the possibility of mechanically amplifying NMR signal~\cite{Sillanpaa2011} and even laser cooling nuclear spins~\cite{Taylor2011,Wood2014,Wood2016} to further enhance the sensitivity of NMR.

\section{Experiment}\label{sec:experimental}

\subsection{Experimental setup}

We aimed at transducing $^{1}$H NMR signals induced in a magnetic field of $\approx 1$~T from the original rf domain ($\omega_{\mathrm{s}}/2\pi \approx 43$~MHz) to the optical domain ($\Omega_{\mathrm{c}}/2 \pi \approx 300$~THz) for a demonstration of EMO NMR.
Figure~\ref{fig:setup} illustrates the experimental setup. For the opto-mechanical and the electro-mechanical couplings, the mechanically compliant part was a high-stress silicon nitride ($\mathrm{Si_{3}N_{4}}$) membrane (Norcada) with lateral dimensions of $0.5 \times 0.5$~mm and a thickness of 50~nm. On the membrane was deposited a circular Au layer with a diameter of 0.45~mm and a thickness of 100~nm. The effective mass $m$ of the Au-coated membrane oscillator was $8.6 \times 10^{-11}$~kg. We found the fundamental $(1,1)$-drum mode oscillation of the Au-coated membrane at $\omega_{\mathrm{m}}/2\pi \approx 180$~kHz. The Q factor was about 1,800 in vacuum with no air damping. Counter electrodes were patterned on a silica plate, and the membrane capacitor was assembled with a designed gap $d_{0}$ between the electrodes of 800 nm. The actual gap was estimatedto be $d_{0} \approx 1.4~\mu$m~(see Appendex for more detail).

The magnetic field was provided by a nominally 1~T permanent magnet, in which a pair of orthogonal rf coils were embedded for pulsed excitation of nuclear spins and NMR signal reception, respectively. The excitation coil was a 2-turn saddle coil, while the detection coil was a 10-turn solenoid coil with a diameter of 3~mm ($L=$ 150~nH). In addition, a pair of planar coils (not shown) were placed outside the rf coil pairs to vary the static magnetic field with application of dc current around the resonance condition of the proton spins.
The membrane capacitor was connected in parallel with the detection coil together with additional trimmer capacitors with capacitances $C_{\mathrm{t}}=98$ pF and $C_{\mathrm{m}}=21$ pF, forming a balanced resonant circuit at $\omega_{\mathrm{LC}}/2 \pi \approx \omega_{\mathrm{s}}/2 \pi \approx$~43~MHz with the Q factor of 26.7. The excitation coil was also impedance-matched at the same frequency. The isolation between these two separate circuits was 22.5~dB at the resonance frequency.

The design of the optical Fabry-P\'{e}rot cavity is described in Appendix. Here, the metal-coated membrane served as one of the two mirrors of an optical cavity for a laser beam with a wavelength of 780~nm. The other mirror with a reflectance of 97\% and a radius of curvature of 75~mm was attached to a ring piezo actuator. The cavity length, which was coarsely adjusted to 17.5~mm, was locked by the feedback on the piezo to the position where the amplitude of the reflected laser beam drops half the dip at cavity resonance, so that the membrane oscillation resulted in amplitude modulation of the laser and thus was imprinted in the optical sideband signal at $\omega_{\mathrm{m}}$.
Note that the cutoff frequency of the piezo servo system is far below $\omega_{\mathrm{m}}$, so that the mechanical resoponse, which would include the rf signal contribution, can be safely transduced to the optical sideband signal at $\omega_{\mathrm{m}}$.

\subsection{Electro-mechano-optical signal transduction} \label{sec.EMOst}

The rf signal developed in the detection LC circuit was parametrically transduced to the membrane oscillation in the presence of the drive signal at either the sum or the difference angular frequency $\omega_{\mathrm{D}} = \omega_{\mathrm{s}} \pm \omega_{\mathrm{m}}$, which was applied to bridge the mismatch between the $^{1}$H resonance frequency $\omega_{\mathrm{s}}/2 \pi \approx 43$~MHz and the membrane resonance frequency $\omega_{\mathrm{m}}/2 \pi \approx 180 $ kHz. The resultant membrane oscillation was then probed by light.

\begin{figure}[t]
\begin{center}
\includegraphics[width=\linewidth]{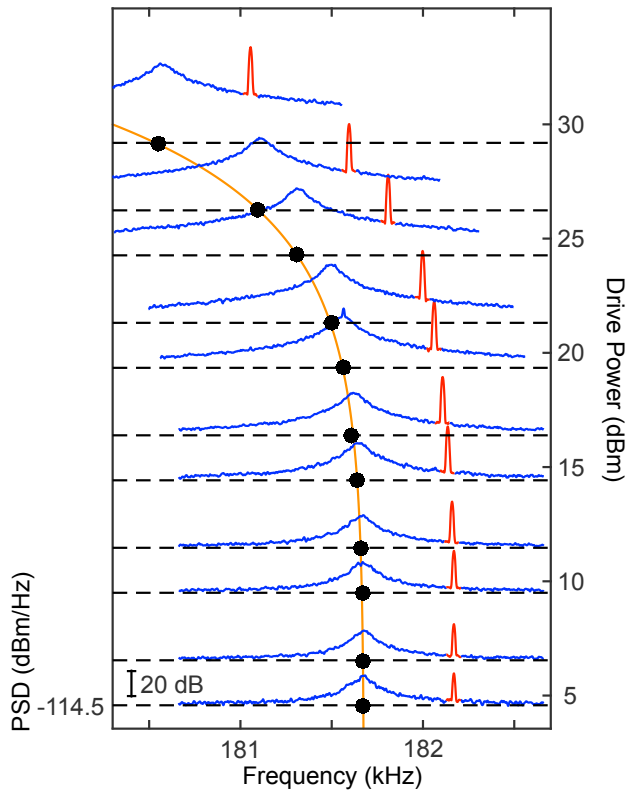}
\caption{Drive-power dependence of the sideband spectra of the optically-detected membrane oscillation under application of a continuous-wave tone signal with a power of $-81$~dBm. The spectra are plotted with vertical offsets proportional to the drive power. The baselines (horizontal broken lines) indicate the corresponding drive power (right axis) as well as the reference power spectral density of $-114.5$~dBm/Hz. Along with the membrane spectra (blue lines) the peaks corresponding to the tone signals (red lines) appear at 500-Hz off-resonance from the mechanical resonance frequency $\omega_{\mathrm{m}}$ (black points). The observed downward shifts of the mechanical resonance frequency were fitted with a model discussed in Appendix (orange line).} \label{fig:extone}
\end{center}
\end{figure}

To examine the EMO signal transduction, we applied to port A in Fig.~\ref{fig:setup} a continuous-wave rf signal at a frequency $\omega_{\mathrm{s}} / 2\pi + 500$~Hz, instead of the real emf signal, together with drive irradiation at various powers. Figure~\ref{fig:extone} shows the acquired optical sideband spectra, where in each spectrum the mechanical responses of the membrane to the noise (blue) as well as the delta-function-like rf-signal tone (red) are visible. With increasing power of the drive, the mechanical resonance frequency is shifted downward~\cite{Polzik2014}. In addition to the Johnson noise and the Brownian noise of the mechanical oscillator, we found increase in the noise floor with the drive power. We ascribed this to the phase noise of the drive as will describe in Sec.~\ref{sec:theory}\ref{sec:comparisonToExp}.

\subsection{$^{1}$H spin echo experiment}

\begin{figure}[htbp]
\begin{center}
\includegraphics[width=\linewidth]{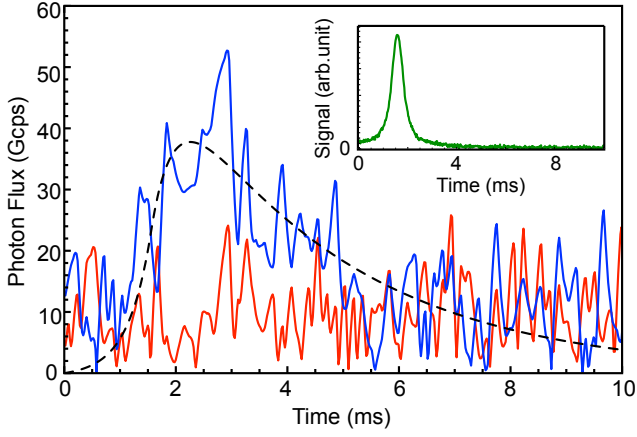}
\caption{$^1$H spin echo signals in 0.1 mol/L aqueous solution of CuSO$_4$ detected by the EMO approach on-resonance (blue line) and $+2.5$-kHz off-resonance (red line). The vertical scale represents the 5000-times average signal intensity in units of the number of photons reaching the photo-detector per second. The broken line represents a convolution of the electrically-detected spin-echo signal shown in the inset with an exponential function with a time constant $2/\gamma_{\mathrm{m}}$. The signal-to-noise ratio $S/N$ is about $5.4$.} \label{fig:omesignal}
\end{center}
\end{figure}

$^{1}$H NMR experiments were then carried out at room temperature using 0.1 mol/L aqueous solution of CuSO$_4$ in a glass test tube (inner diameter 1 mm) with $\approx 2.2\times 10^{20}$ $^{1}$H spins of water molecules, in which the paramagnetic copper ions accelerate $^{1}$H spin relaxation, allowing rapid repetition of signal averaging. The spin-echo measurement~\cite{Hahn1950} was performed by applying rf pulses with a power of +17 dBm to the tuned excitation coil through port~B in Fig.~\ref{fig:setup} with the widths of the $\pi/2$ and the $\pi$ pulses of 140 $\mu$s and 280 $\mu$s, respectively, and the pulse interval of 1.5 ms. The inset of Fig.~\ref{fig:omesignal} shows a conventional electrical signal of the $^1$H spin echo obtained by connecting port~A in Fig.~\ref{fig:setup} to a low-noise amplifier, so that the amplified electrical nuclear induction signal could be sent to the conventional demodulation circuit of the NMR spectrometer.
The maximum intensity of the NMR echo signal was $-93$~dBm at the input of the low-noise amplifier. The observed decay with a time constant $T_{2}^{*} \approx 320$~$\mu$s was dominantly caused by the inhomogeneity of the magnetic field.

Next, the low-noise amplifier at port~A in Fig.~\ref{fig:setup} was replaced with a drive source for down conversion of the NMR signal to the mechanical frequency, and the optical output from the Fabry-P\'{e}rot cavity was measured under the drive power of +15~dBm~. During the rf pulses, the frequency of the drive was detuned by +400~kHz, so as to decouple the electro-mechanical interaction and thereby prevent the membrane from being shaken by the excitation rf pulse leaked to the detection circuit, which, in spite of the 22.5 dB isolation, was still orders of magnitude more intense than the NMR signals induced in the receiving LC circuit ($-93$~dBm).

Figure~\ref{fig:omesignal} shows the electro-mechano-optically detected spin-echo signal (blue line) accumulated over 5000 times with a repetition interval of 20 ms. For comparison, we performed another measurement with the identical experimental parameters except for a slight shift in the static magnetic field ($\approx$0.06 mT) to make the $^1$H spins off-resonant by 2.5 kHz, and verified that the signal disappeared (red line), convincing ourselves that the profile of the optically detected signal (blue line in Fig.~\ref{fig:omesignal}) does really originate from the nuclear induction signal.

The difference in the profile of the spin-echo signal obtained by the EMO approach from that in the conventional electrical scheme can be explained by the transient response of the high-Q membrane. That is, the response $b(t)$ of the membrane to an excitation $a(t)$, the present case of which is the profile of the electrically detected spin echo, is determined by the response function $h(t)$ of the membrane through convolution, i.e., $b(t) = \int_{-\infty}^{t} h(t-\tau) * a(\tau) d \tau$. Since the spectrum of the fundamental mode of the membrane was well fitted with a Lorentzian function with a width $\gamma_{\mathrm{m}}/2 \pi \approx 100$~Hz, we approximated the response function $h(t)$ to be an exponentially decaying function with a time constant $2/\gamma_{\mathrm{m}}$, and calculated the response $b(t)$, which was found to reproduce the measured profile of the EMO NMR signal (broken line in Fig.~\ref{fig:omesignal}).

\begin{figure*} [htbp]
\begin{center}
\includegraphics[width=\linewidth]{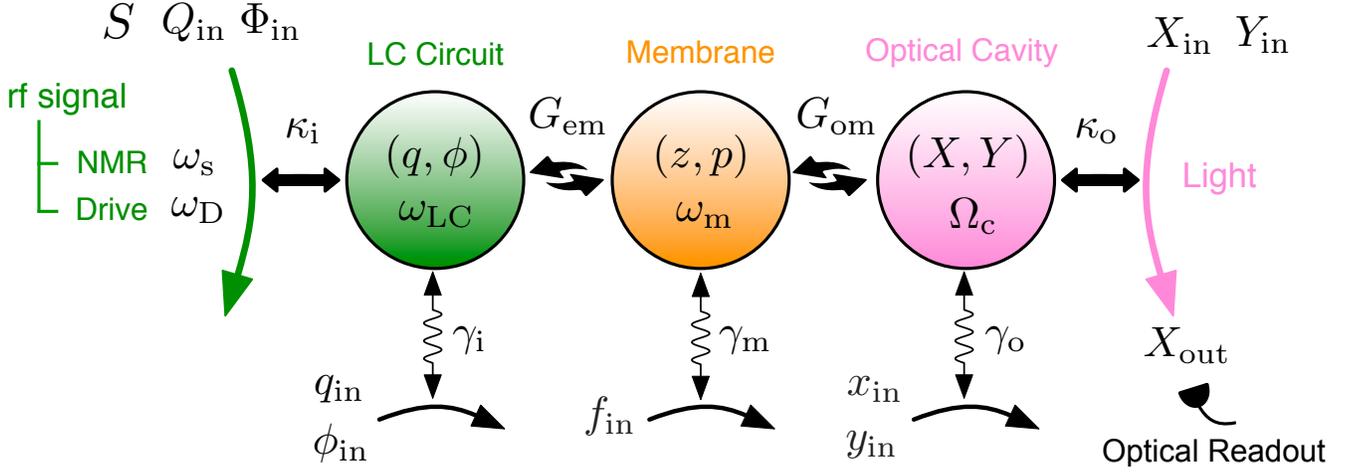}
\caption{Schematic diagram of electro-mechano-optical signal transduction of NMR. The three harmonic oscillators, namely, the LC circuit, the membrane, and the optical cavity, are represented with circles, each of which has channels of the input and output, with coupling strengths $\kappa_{\mathrm{i}}, G_{\mathrm{em}}, G_{\mathrm{om}}$, and $\kappa_{\mathrm{o}}$, and dissipation to the bath, with rates $\gamma_{\mathrm{i}}, \gamma_{\mathrm{m}}$, and $\gamma_{\mathrm{o}}$. The rf signal $S$ generated by nuclear induction at frequency $\omega_{\mathrm{s}} \approx \omega_{\mathrm{LC}}$ is transduced to the membrane oscillation through the LC circuit with the electro-mechanical coupling under application of the drive signal at $\omega_{\mathrm{D}}=\omega_{\mathrm{LC}} + \omega_{\mathrm{m}}$. The resultant membrane oscillation is in turn read out optically with the optical cavity through the opto-mechanical coupling.} \label{fig:scheme}
\end{center}
\end{figure*}

\section{Theory and discussion}\label{sec:theory}
\subsection{Dynamics of the EMO system}
Figure~\ref{fig:scheme} schematically shows the pathway of successive signal transduction through a chain of three harmonic oscillators, namely, the LC circuit, the membrane oscillator, and the optical cavity.
Here, $q$ and $\phi$ are the charge and the flux of the LC circuit, $z$ and $p$ are the displacement and the momentum of the mechanical oscillator, and $X$ and $Y$ are the canonical quadratures of the optical cavity's field. $G_{\mathrm{em}}$ and $G_{\mathrm{om}}$ are the electro-mechanical and the opto-mechanical coupling strength.
$\gamma_{\mathrm{i}}$, $\gamma_{\mathrm{m}}$, and $\gamma_{\mathrm{o}}$ are the intrinsic dissipation rates for the LC circuit, the mechanical oscillator, and the optical cavity. Associated with these dissipations there are \textit{rotating-frame} thermal fluctuation inputs $q_{\mathrm{in}}$ and $\phi_{\mathrm{in}}$ for the LC circuit, \textit{laboratory-frame} thermal fluctuation input $f_{\mathrm{in}}$ for the mechanical oscillator. The thermal fluctuation inputs for the optical cavity, $x_{\mathrm{in}}$ and $y_{\mathrm{in}}$, are negligible and thus omitted. $\kappa_{\mathrm{i}}$ and $\kappa_{\mathrm{o}}$ are the external coupling rates for the LC circuit and the optical cavity, and, in addition to the NMR signal input, $S$, the associated fluctuation inputs are $Q_{\mathrm{in}}$ and $\Phi_{\mathrm{in}}$ for the LC circuit and $X_{\mathrm{in}}$ and $Y_{\mathrm{in}}$ for the optical cavity. The total dissipation rates are thus $\kappa_{\mathrm{iT}} = \kappa_{\mathrm{i}}+\gamma_{\mathrm{i}}$ for the LC circuit and $\kappa_{\mathrm{oT}}=\kappa_{\mathrm{o}}+\gamma_{\mathrm{o}}$ for the optical cavity, respectively.

Using the input-output formalism in the rotating wave approximation~\cite{Clerk2010}, 
we have the following Heisenberg-Langevin equations of motion:
\begin{align}
\dot{q} &= -\Delta_{\mathrm{i}} \phi - \frac{\kappa_{\mathrm{iT}}}{2} q - \sqrt{\kappa_{\mathrm{i}}} (Q_{\mathrm{in}} + S ) - \sqrt{\gamma_{\mathrm{i}}} q_{\mathrm{in}} \label{eq:q} \\
\dot{\phi} &= \Delta_{\mathrm{i}} q - \frac{\kappa_{\mathrm{iT}}}{2} \phi - \sqrt{\kappa_{\mathrm{i}}} \Phi_{\mathrm{in}} - \sqrt{\gamma_{\mathrm{i}}} \phi_{\mathrm{in}} -G_{\mathrm{em}} z  \label{eq:phi} \\
\dot{z} &= \omega_{\mathrm{m}} p \label{eq:x0} \\
\dot{p} &= -\omega_{\mathrm{m}} z - \gamma_{\mathrm{m}} p - \sqrt{2 \gamma_{\mathrm{m}}} f_{\mathrm{in}} -G_{\mathrm{em}} q - G_{\mathrm{om}} X  \label{eq:p0} \\
\dot{X} &= -\Delta_{\mathrm{o}} Y - \frac{\kappa_{\mathrm{oT}}}{2} X - \sqrt{\kappa_{\mathrm{o}}} X_{\mathrm{in}} \label{eq:X} \\
\dot{Y} &= \Delta_{\mathrm{o}} X - \frac{\kappa_{\mathrm{oT}}}{2} Y  -\sqrt{\kappa_{\mathrm{o}}} Y_{\mathrm{in}} -G_{\mathrm{om}}z, \label{eq:Y}
\end{align}
where $\Delta_{\mathrm{i}} = \omega_{\mathrm{D}}-\omega_{\mathrm{LC}}$ is the difference between the drive signal frequency and the LC resonance frequency, and $\Delta_{\mathrm{o}}=\Omega_{\mathrm{D}}-\Omega_{\mathrm{c}}$ is the detuning of the optical cavity from the drive laser frequency. We note that the electro-mechanical coupling $G_{\mathrm{em}}$ increases quadratically with decreasing the gap $d_{0}$ between the electrodes of the capacitor~(see Appendix).

Now we shall see how the rf NMR signal, $S$ appearing in Eq.~(\ref{eq:q}), is transduced to the optical output $X_{\mathrm{out}}$, which is given by the input-output relation
\begin{align}
  X_{\mathrm{out}} = X_{\mathrm{in}} + \sqrt{\kappa_{\mathrm{o}}} X.
\end{align}
By taking the time derivative of Eq.~(\ref{eq:x0}) and using Eq.~(\ref{eq:p0}) we have
\begin{equation}
\frac{\ddot{z}}{\omega_{\mathrm{m}}} = - \omega_{\mathrm{m}} z - \frac{\gamma_{\mathrm{m}}}{\omega_{\mathrm{m}}} \dot{z} -\sqrt{2 \gamma_{\mathrm{m}}} f_{\mathrm{in}} -\left( G_{\mathrm{em}} q + G_{\mathrm{om}} X \right), \label{eq:x}
\end{equation}
where $f_{\mathrm{in}} \equiv p_{\mathrm{in}}+ \dot{z}_{in}/\omega_{0}$ is the mechanical thermal noise input.
In the frequency domain, the above equation for the displacement can be written as
\begin{equation}
z (\omega) = \chi_{\mathrm{m}}(\omega) \left[- \sqrt{2 \gamma_{\mathrm{m}}} f_{\mathrm{in}}(\omega) - \left( G_{\mathrm{em}} q(\omega) + G_{\mathrm{om}}X(\omega) \right) \right], \label{eq:xF}
\end{equation}
where the mechanical susceptibility $\chi_{\mathrm{m}}(\omega)$ is defined by
\begin{equation}
\chi_{\mathrm{m}} (\omega) =
\left( -\frac{\omega^{2}}{\omega_{\mathrm{m}}} -i \frac{\omega \gamma_{\mathrm{m}}}{\omega_{\mathrm{m}}} + \omega_{\mathrm{m}} \right)^{-1}.
\end{equation}
In a similar fashion, we obtain the frequency-domain representaion for $q$ and $X$ as
\begin{widetext}
  \begin{align}
    q(\omega) &= \chi_{\mathrm{LC}}(\omega) 
    \left[ \Delta_{\mathrm{i}}G_{\mathrm{em}} z + \sqrt{\kappa_{\mathrm{i}}} \left( \Delta_{\mathrm{i}} \Phi_{\mathrm{in}} + \left(i \omega - \frac{\kappa_{\mathrm{iT}}}{2} \right) Q_{\mathrm{in}} \right)
    + \sqrt{\gamma_{\mathrm{i}}} \left( \Delta_{\mathrm{i}} \phi_{\mathrm{in}} + \left(i \omega - \frac{\kappa_{\mathrm{iT}}}{2} \right) q_{\mathrm{in}} \right) \right],  \label{eq:q0} \\
  X(\omega) &= \chi_{\mathrm{c}}(\omega) \left[ \Delta_{\mathrm{o}}G_{\mathrm{om}} z + \sqrt{\kappa_{\mathrm{o}}} \left( \Delta_{\mathrm{o}} Y_{\mathrm{in}} + \left(i \omega - \frac{\kappa_{\mathrm{oT}}}{2} \right) X_{\mathrm{in}} \right) \right], \label{eq:X0}
  \end{align}
\end{widetext}
where the LC susceptibility $\chi_{\mathrm{LC}}(\omega)$ and the optical susceptibility $\chi_{\mathrm{c}}(\omega)$ are given by
\begin{align}
  \chi_{\mathrm{LC}}(\omega) &= \left[ \left(-i \omega + \frac{\kappa_{\mathrm{iT}}}{2} \right)^{2} + \Delta_{\mathrm{i}}^{2} \right]^{-1}, \\
  \chi_{\mathrm{c}}(\omega) &= \left[ \left(-i \omega + \frac{\kappa_{\mathrm{oT}}}{2} \right)^{2} + \Delta_{\mathrm{o}}^{2} \right]^{-1}.\label{eq:chic}
\end{align}

In our experiment, $\Delta_{\mathrm{i}} \approx \omega_{\mathrm{m}}$ was much smaller than the resonant bandwidth of the LC circuit. Thus, we consider the case of resonant application of the drive, $\Delta_{\mathrm{i}} \rightarrow 0$.
In addition, to detect the membrane displacement through amplitude modulation of the optical output, we detuned the optical cavity approximately by half its bandwidth, i.e., $\Delta_{\mathrm{o}} \approx \kappa_{\mathrm{oT}}/2$.
Further, since the frequency $\omega_{\mathrm{m}}$ of interest in the optical output signal is much smaller than $\Delta_{\mathrm{o}}$, we set $\omega \rightarrow 0$ in Eqs.~(\ref{eq:X0}) and (\ref{eq:chic}).
Neglecting $G_{\mathrm{om}}^2 \ll 1$, we obtain after some algebra
\begin{widetext}
\begin{align}
X_{\mathrm{out}} =  \frac{\kappa_{\mathrm{o}}}{\kappa_{\mathrm{oT}}}Y_{\mathrm{in}} + \left( 1-\frac{\kappa_{\mathrm{o}}}{\kappa_{\mathrm{oT}}} \right) X_{\mathrm{in}} - \frac{G_{\mathrm{om}} \sqrt{\kappa_{\mathrm{o}}}}{\kappa_{\mathrm{oT}}} \sqrt{2\gamma_{\mathrm{m}}} \chi_{\mathrm{m}}(\omega) f_{\mathrm{in}}
 - \frac{G_{\mathrm{om}}\sqrt{\kappa_{\mathrm{o}}}}{\kappa_{\mathrm{oT}}} \frac{G_{\mathrm{em}}}{i \omega-\frac{\kappa_{\mathrm{iT}}}{2}}\chi_{\mathrm{m}}(\omega) \left[ \sqrt{\kappa_{\mathrm{i}}} \left( Q_{\mathrm{in}} + S \right) + \sqrt{\gamma_{\mathrm{i}}} q_{\mathrm{in}} \right]. \label{eq:Xo}
\end{align}
In the \textit{laboratory frame}, the linearized rotating-frame signal $X_{\mathrm{out}}$ in Eq.~(\ref{eq:Xo}) has to be modified to be
\begin{equation}
\tilde{X}_{\mathrm{out}}=X_{\mathrm{out}} \cos (\Omega t) + Y_{\mathrm{out}} \sin (\Omega t) + \sqrt{\kappa_{\mathrm{o}} \mathcal{N}_{\mathrm{D}}} \cos (\Omega_{\mathrm{D}}t)
\end{equation}
where the last term comes from the \textit{displacement} by the optical drive, which oscillates at frequency $\Omega_{\mathrm{D}}$. Here, $\mathcal{N}_{\mathrm{D}}$ is the intracavity photon number (see Appendix). Note that $Y_{\mathrm{out}}$, now appearing in $\tilde{X}_{\mathrm{out}}$, is given by
\begin{align}
Y_{\mathrm{out}} =  -\frac{\kappa_{\mathrm{o}}}{\kappa_{\mathrm{oT}}}X_{\mathrm{in}} + \left( 1-\frac{\kappa_{\mathrm{o}}}{\kappa_{\mathrm{oT}}} \right) Y_{\mathrm{in}} + \frac{G_{\mathrm{om}} \sqrt{\kappa_{\mathrm{o}}}}{\kappa_{\mathrm{oT}}} \sqrt{2\gamma_{\mathrm{m}}} \chi_{\mathrm{m}}(\omega) f_{\mathrm{in}}
+ \frac{G_{\mathrm{om}}\sqrt{\kappa_{\mathrm{o}}}}{\kappa_{\mathrm{oT}}} \frac{G_{\mathrm{em}}}{i \omega-\frac{\kappa_{\mathrm{iT}}}{2}}\chi_{\mathrm{m}}(\omega) \left[ \sqrt{\kappa_{\mathrm{i}}} \left( Q_{\mathrm{in}} + S \right) + \sqrt{\gamma_{\mathrm{i}}} q_{\mathrm{in}} \right]. \label{eq:Yo}
\end{align}
\end{widetext}
In the photo-detected signal $\left|\tilde{X}_{\mathrm{out}}\right|^{2}$ in the laboratory frame, the components oscillating around $\omega \sim \omega_{\mathrm{m}}$, which are produced by the interference between the term oscillating at $\Omega_{\mathrm{D}}$ and the ones at $\Omega_{\mathrm{D}} \pm \omega_{\mathrm{m}}$, are of interest. These components constitute the \textit{optical signal output}, $O(\omega)$, which amounts to the magnitude of the quadrature demodulated signal (see Appendix) and can be written as
\begin{equation}
O(\omega) = \sqrt{\kappa_{\mathrm{o}} \mathcal{N}_{\mathrm{D}}} \sqrt{\left| X_{\mathrm{out}} \right|^{2} + \left|Y_{\mathrm{out}} \right|^{2}}. \label{eq:Xo2}
\end{equation}
This indeed contains the \textit{rf signal input} $S$ along with various noises, which is faithfully transduced from the mechanical response Eq.~(\ref{eq:xF}) with the amplification factor proportional to $G_{\mathrm{om}}$ as seen in Eqs.~(\ref{eq:Xo}) and (\ref{eq:Yo}). The added noise here is just the optical shot noise, which can be quantum-noise-limited. One of the potential advantage of the EMO NMR detection over the convential NMR is thus the fact that both the Brownian noise and the optical shot noise can be suppressed by increasing the electro-mechanical coupling $G_{\mathrm{em}}$ as well as the opto-mechanical coupling $G_{\mathrm{om}}$~\cite{Polzik2014}.

\subsection{Noise spectral densities}
Since the mean value of noise is zero, each noise shall be evaluated in terms of spectral density.
For the Brownian noise of the mechanical oscillator, the noise spectral density $S_{FF}$ is defined as $S_{FF}= \left| f_{\mathrm{in}} \right|^{2}$.
The Johnson noise in the LC circuit can come from the bath as well as from the input channel, and its spectral density, $S_{qq}$, is given $\kappa_{\mathrm{iT}} S_{qq} = \kappa_{\mathrm{i}}\left| Q_{\mathrm{in}} \right|^{2} + \gamma_{\mathrm{i}}\left| q_{\mathrm{in}} \right|^{2}$.
Assuming that these noise spectra $S_{FF}(\omega)$ and $S_{qq}(\omega)$ are white within the bandwidth of the mechanical resonance, we have the Nyquist-type noise spectra,
\begin{align}
S_{FF}(\omega) &= n_{\mathrm{th}}(\omega_{\mathrm{m}},T_{\mathrm{eff}}), \label{eq:ntm}\\
S_{qq}(\omega) &= n_{\mathrm{th}}(\omega_{\mathrm{LC}},T), \label{eq:ntLC}
\end{align}
with
\begin{align}
  n_{\mathrm{th}}(\omega,T) = \frac{k_{\mathrm{B}}T}{\hbar \omega}.
\end{align}
Here, we assume that the \textit{electric bath temperature} $T$ is 300~K, while the \textit{mechanical bath temperature} $T_{\mathrm{eff}}$ is not necessarily equal to 300~K but can rather be higher given that the quality factor is good so that the ambient noise could easily bring the mechanical oscillator away from the thermal equilibrium. We note that the LC circuit and the mechanical oscillator are both in a high temperarure regime where $k_{\mathrm{B}}T_{\mathrm{eff}} \gg \hbar \omega_{\mathrm{m}}$ and $k_{\mathrm{B}} T \gg \hbar \omega_{\mathrm{LC}}$.
Conversely, we can expect that the noise spectral density $S_{XX} \equiv | X_{\mathrm{in}} |^{2}$ and $S_{YY} \equiv | Y_{\mathrm{in}} |^{2}$ for the optical part can be made much smaller.

From Eq.~(\ref{eq:Xo2}), the single-sided spectral density $S_{\mathrm{oo}}(\omega)$ of the optical signal at frequency $\omega$ close to $\omega_{\mathrm{m}}$ can be written as
\begin{widetext}
\begin{align}
S_{\mathrm{oo}}(\omega) =& \kappa_{\mathrm{o}} \mathcal{N}_{\mathrm{D}} \left[ \left( \left(\frac{\kappa_{\mathrm{o}}}{\kappa_{\mathrm{oT}}} \right)^{2} +  \left(1-\frac{\kappa_{\mathrm{o}}}{\kappa_{\mathrm{oT}}} \right)^{2} \right) \left( 2S_{XX}(\omega) + 2S_{YY}(\omega) \right) \right.
 + C_{\mathrm{om}}\frac{\kappa_{\mathrm{o}}}{\kappa_{\mathrm{oT}}} 2 \gamma_{\mathrm{m}}^{2} \left| \chi_{\mathrm{m}}(\omega) \right|^{2} 4S_{FF}(\omega) \nonumber \\
& + C_{\mathrm{om}}\frac{\kappa_{\mathrm{o}}}{\kappa_{\mathrm{oT}}} C_{\mathrm{em}}(\omega) \gamma_{\mathrm{m}}^{2} \left| \chi_{\mathrm{m}}(\omega) \right|^{2}
 \left.
  \left[ 4 S_{qq}(\omega) + \frac{\kappa_{\mathrm{i}}}{\kappa_{\mathrm{iT}}} 4S^{2} \delta \left( \omega-\omega_{\mathrm{m}} \right) \right] \right]. \label{eq:Soo}
\end{align}
\end{widetext}
Here, we introduced the \textit{opto-mechanical cooperativity} $C_{\mathrm{om}}$ and the frequency-dependent \textit{electro-mechanical cooperativity} $C_{\mathrm{em}}(\omega)$ as
\begin{align}
C_{\mathrm{om}} &= \frac{G_{\mathrm{om}}^{2}}{\gamma_{\mathrm{m}} \kappa_{\mathrm{oT}}}, \label{eq:Com} \\
C_{\mathrm{em}}(\omega) &= \frac{4 G_{\mathrm{em}}^{2}}{\gamma_{\mathrm{m}} \kappa_{\mathrm{iT}}}\frac{\kappa_{\mathrm{iT}}^{2}}{4 \omega^{2} + \kappa_{\mathrm{iT}}^{2}}. \label{eq:Cem}
\end{align}

\subsection{Signal-to-noise ratio}
In the under-coupling limit $\kappa_{\mathrm{o}} \ll \kappa_{\mathrm{oT}}$, the signal-to-noise ratio $S/N$ in units of photon number within a narrow frequency range $\Delta \ll \omega_{\mathrm{m}}$ at around $\omega=\omega_{\mathrm{m}}$, i.e., from $\omega_{\mathrm{m}}-\frac{\Delta}{2}$ to $\omega_{\mathrm{m}}+\frac{\Delta}{2}$, is
\begin{widetext}
\begin{align}
 \sqrt{\frac{S^{2}}{\cfrac{\kappa_{\mathrm{iT}}}{\kappa_{\mathrm{i}}} \left( \cfrac{S_{XX}(\omega_{\mathrm{m}}) + S_{YY}(\omega_{\mathrm{m}})}{2 C_{\mathrm{om}} \frac{\kappa_{\mathrm{o}}}{\kappa_{\mathrm{oT}}} C_{\mathrm{em}}(\omega_{\mathrm{m}}) } + \cfrac{2 S_{FF}(\omega_{\mathrm{m}})}{C_{\mathrm{em}}(\omega_{\mathrm{m}})} + S_{qq}(\omega_{\mathrm{m}}) \right) \Delta}}, \label{eq:SNR}
\end{align}
\end{widetext}
where we used $\gamma_{\mathrm{m}}^{2} \left| \chi_{\mathrm{m}}(\omega_{\mathrm{m}}) \right|^{2}=1$. The form of the signal-to-noise ratio consolidates the aforementioned potential advantege of the EMO NMR. All the noise except for the Johnson noise, which is intrinsically inseperable from the rf signal, are suppressed by increasing the electro-mechanical coupling $G_{\mathrm{em}}$ and thus the electro-mechanical cooperativity $C_{\mathrm{em}}$~\cite{Polzik2014}.

Note that the yet another figure-of-merit, \textit{signal tranfer rate}~\cite{Zeuthen2016}, for the current EMO-NMR is given by $C_{\mathrm{om}} \frac{\kappa_{\mathrm{o}}}{\kappa_{\mathrm{oT}}} C_{\mathrm{em}} \frac{\kappa_{\mathrm{i}}}{\kappa_{\mathrm{iT}}}$.

\begin{table*}[t]
 \caption{Noise budget of the current EMO NMR detection.}
  \begin{center}
   \begin{tabular}{|c||c|c|c|c||c|}
    \hline
      \ & Shot noise & Brownian noise & Johnson noise & Phase noise & Total noise \\
    \hline \hline
     Symbolic notation$^{\cfrac{}{}}_{\cfrac{}{}}$ & $\ \cfrac{\kappa_{\mathrm{iT}}}{\kappa_{\mathrm{i}}} \cfrac{S_{XX}+S_{YY}}{2 C_{\mathrm{om}} \frac{\kappa_{\mathrm{o}}}{\kappa_{\mathrm{oT}}} C_{\mathrm{em}}} \ $ & $2\cfrac{\kappa_{\mathrm{iT}}}{\kappa_{\mathrm{i}}} \cfrac{S_{FF}}{C_{\mathrm{em}}}$ &  $\cfrac{\kappa_{\mathrm{iT}}}{\kappa_{\mathrm{i}}} S_{qq}$ & $\eta_{\mathrm{p}} \cfrac{P_{\mathrm{D}}}{\hbar \omega_{\mathrm{D}} \gamma_{\mathrm{m}}}$ &  \\
    \hline
      Number of quanta & $4.4 \times 10^6$ & $5.0 \times 10^9$ & $3.3 \times 10^5$ & $1.9 \times 10^{10}$ & $2.4 \times 10^{10}$ \\
    \hline
      Effective temperature [K] & $ 8000 $ & $9.2 \times 10^6$ & 590 & $3.5 \times 10^7$ & $4.4 \times 10^7$ \\
    \hline
   \end{tabular}
  \end{center} \label{tb:c}
\end{table*}

\subsection{Comparison to the experiments} \label{sec:comparisonToExp}
We calibrated the parameters~(see Appendix) that characterize the EMO signal transduction from the acquired optical sideband spectra shown in Fig.~\ref{fig:extone}. In the presence of +15~dBm drive, the electro-mechanical cooperativity $C_{\mathrm{em}}$ of 0.019 was attained, whereas the opto-mechanical cooperativity $C_{\mathrm{om}}$ was 0.32~$\times 10^{-3}$. With these values the signal transfer rate amounts to $\sim 1.1 \times 10^{-7}$.

As the drive power is increased to make $C_{\mathrm{em}}$ much larger, however, the phase noise of the drive becomes conspicuous as mentioned in Sec.~\ref{sec:experimental}\ref{sec.EMOst}.
In terms of the offset angular frequency $\omega$ the profile of the phase noise can be expressed as,
\begin{equation}
\mathcal{L}(\omega) = \frac{\delta_{\mathrm{P}}}{\omega^{2} + \frac{\delta_{\mathrm{P}}^{\ 2}}{4}}, \label{eq:L}
\end{equation}
a Lorentzian form with the spectral line width of $\delta_{\mathrm{P}}$, where $1/f$ noise and frequency-independent noise are ignored. Then the photon flux associated with the phase noise of the drive at sideband frequency $\omega$ can be given by
\begin{equation}
\mathcal{L}(\omega) \frac{P_{\mathrm{D}}}{\hbar \omega_{\mathrm{D}}},
\end{equation}
where $P_{\mathrm{D}}$ is the  power of the drive.
Thus, the spectral density at frequency $\omega$ in Eq.~(\ref{eq:Soo}) needs to be modified when the phase-noise contribution is appreciable.

To deduce the expected signal-to-noise ratio, one missing element is the bandwidth of the NMR signal. In the echo experiment, the effective bandwidth of the detection is determined by $1/\pi T_{2}^{*} \approx 1$~kHz where $T_{2}^{*} \approx 320$~$\mu$s. Since the bandwidth of the electro-mechano-optical NMR detection is limited by the mechanical response, $\Delta/2 \pi \approxeq \gamma_{\mathrm{m}}/2 \pi \approx 100$~Hz, the impedance mismatch roughly leads to the factor of $\gamma_{\mathrm{m}} T_{2}^{*} / 2$ reduction of the signal strength. The signal-to-noise ratio for the echo experiment shown in Fig.~4 is thus expected to be
\begin{widetext}
\begin{equation}
\frac{S}{N} = \sqrt{\frac{S^{2} \frac{T_{2}^{*}}{2} \left( \gamma_{\mathrm{m}} \frac{T_{2}^{*}}{2} \right) }{\cfrac{\kappa_{\mathrm{iT}}}{\kappa_{\mathrm{i}}} \cfrac{S_{XX}(\omega_{\mathrm{m}})+S_{YY}(\omega_{\mathrm{m}})}{2 C_{\mathrm{om}} \frac{\kappa_{\mathrm{o}}}{\kappa_{\mathrm{oT}}} C_{\mathrm{em}}(\omega_{\mathrm{m}}) } + \cfrac{\kappa_{\mathrm{iT}}}{\kappa_{\mathrm{i}}} \cfrac{2 S_{FF}(\omega_{\mathrm{m}})}{C_{\mathrm{em}}(\omega_{\mathrm{m}})} + \cfrac{\kappa_{\mathrm{iT}}}{\kappa_{\mathrm{i}}} S_{qq}(\omega_{\mathrm{m}}) + \eta_{\mathrm{p}} \cfrac{P_{\mathrm{D}}}{\hbar \omega_{\mathrm{D}} \gamma_{\mathrm{m}}}}} \approx 0.12 \label{eq:SNR4}
\end{equation}
\end{widetext}
for the single-shot measurement, where the parameter $\eta_{\mathrm{p}}$ in Eq.~(\ref{eq:SNR4}) characterizing the phase noise at around $\omega_{\mathrm{m}}$ with respect to the carrier at $\omega_{\mathrm{D}}$, i.e.,
\begin{equation}
\eta_{p} = \int_{\omega_{\mathrm{m}}-\frac{\Delta}{2}}^{\omega_{\mathrm{m}}+\frac{\Delta}{2}} \frac{d \omega}{2 \pi} \gamma_{\mathrm{m}}^{2} \left| \chi_{\mathrm{m}}(\omega) \right|^{2} \mathcal{L}(\omega).
\end{equation}
which was evaluated to be $\eta_{\mathrm{p}} \approx 10^{-11}$ (see Appendix). The number of total noise quanta [the denominator of Eq.~(\ref{eq:SNR4})] is estimated to be $2.4 \times 10^{10}$ while the signal quanta [the numerator of Eq.~(\ref{eq:SNR4})] for the echo experiment is on the order of $3.6 \times 10^{8}$, which are proportional to the noise and the signal voltages squared, respectively. The noise budget of the current EMO NMR detection is shown in Table~\ref{tb:c}. With 5000-times averaging, the signal-to-noise ratio becomes roughly 8, agreeing well with the signal-to-noise ratio of the acquired data ($S/N \approx 5.4$) shown in Fig.~\ref{fig:omesignal}.

\subsection{Prospects}
Even though the signal-to-noise ratio in the present proof-of-principle EMO NMR demonstration is lower than that in the conventional electrical NMR approach, there is plenty of room for improving the sensitivity. In particular, the electro-mechanical cooperativity $C_{\mathrm{em}} \propto 1/ d_{0}^{4}$ would increase dramatically by reducing the capacitor gap.
With a realistic revision including the capacitor design, we have a prospect of attaining the effective noise temperature of as low as 6~K at room temperature operation of the transducer with a $+30$-dBm drive~(see Appendix), which would outperform the conventional NMR approach. If the membrane is put in a cryogenic environment, further improvement is expected.

In addition, the effect of the phase noise of the drive can be made negligibly small by increasing the mechanical oscillation frequency and thereby the difference $\omega_{\mathrm{D}} - \omega_{\mathrm{s}}$. One way to do this would be to reduce the weight of the metal layer deposited on the membrane. Some filters can also be arranged to prevent the phase noise of the drive from exciting the mechanical oscillator.

Moreover, as increasing the electro-mechanical cooperativity $C_{\mathrm{em}}$, signal transduction would be accompanied by parametric signal amplification. So far, in NMR and MRI, parametric amplification has been realized using an LC circuit containing a varactor diode, whose capacitance can be varied electrically~\cite{Qian2012,Qian2013}. The present work would lead to electro-mechanical parametric amplification of NMR/MRI signals.

The usage of the Fabry-P\'{e}rot optical cavity in this work opens the possibility of exploiting the effect of radiation-pressure cooling~\cite{Gigan2006,Arcizet2006,Schliesser2006}. If the opto-mechanical and electro-mechanical couplings as well as the laser power are large enough, the membrane's oscillation modes, and thereby the eigenmode of the LC circuit, can be cooled~\cite{Taylor2011}, implying the possibility of cooling nuclear spins through electro-mechanical and opto-mechanical couplings without physically lowering the temperature of the experimental system.  If the expected challenges, such as the insufficient Q factor and finite dissipation rates to the bath, have been overcome, laser cooing of the nuclear spins would provide a way toward further enhancing the NMR sensitivity. It is worth noting that nuclear-spin laser cooling would not require doping of paramagnetic impurities in the sample of interest, in contrast to the current dynamic nuclear polarization schemes~\cite{Griffin1993}.

In the coupling between a microwave cavity and an ensemble electron spins~\cite{Schuster2010,Kubo2010,Abe2011}, analogous population exchange has been theoretically proposed~\cite{Wood2014,Wood2016} and experimentally reported~\cite{Bienfait2016,Eichler2017}. Its extension to nuclear spins with a cold mechanical nano-resonator is also suggested~\cite{Butler2011}.

With the separate coil used for rf excitation, pulsed-NMR techniques for coherent manipulation of nuclear spin interactions can be applied straightforwardly~\cite{Mehring}, whereas in the receiving part, the bandwidth is limited by that of the membrane oscillator ($\approx 100$ Hz). This can be rather narrow compared to the spectral width of interest in NMR analysis, where the resonance lines can spread due to broadening and/or distribution of isotropic shifts. In this context, the EMO approach is compatible with traditional continuous-wave NMR~\cite{Abragam} as well as recently reported field-sweep NMR~\cite{Takeda2012,Yamada2015}, where the frequency of interest is fixed throughout measurement and the external magnetic field is varied instead.
It is also worth noting that the aforementioned enhancement of the electro-mechanical coupling would cause damping of the membrane's oscillation, and thereby increase the accessible bandwidth.

\section{Summary}
Rf signals of nuclear induction can be up-converted to light through the membrane oscillator that forms a part of both the LC resonant circuit and the optical cavity. The EMO NMR approach presented here potentially offers better sensitivity than that of the conventional electrical detection scheme.

\section*{Funding Information}
Japan Science and Technology Agency (JST) SENTAN (Grant No. 14537844); Japan Science and Technology Agency (JST) ERATO (Grant No. JPMJER1601).

\section*{Acknowledgments}

We are grateful to Y.~Tabuchi, M.~Okada, Y.~Tominaga, M.~Negoro, T.~Koshi, K.~Yamada, M.~Takahashi, A.~Saitoh, K.~Kusuyama, M.~Ataka, H.~Fujita, K.~Lehnert, E.~Zeuthen, A.~S{\o}rensen, A.~Schliesser, and E.~S.~Polzik for fruitful discussions and collaborations.

\appendix

\setcounter{equation}{0}
\setcounter{figure}{0}
\renewcommand{\theequation}{A\arabic{equation}}
\renewcommand{\thefigure}{A\arabic{figure}}
\renewcommand{\thetable}{A\arabic{table}}

\section{Appendix}
\section{Membrane capacitor fabrication} \label{sec:membrane}
A Si-frame supported stoichiometric Si$_{3}$N$_{4}$ membrane was purchased from Norcada (Part Number: QX5050AS). On the membrane with a lateral size of $0.5 \times 0.5$ mm and a thickness of 50 nm, an Au layer was deposited and photo-lithographically patterned with wet-etching into a circular pad with a diameter of 0.45~mm and a thickness of 100~nm [Fig.~1(c)].

\begin{figure} [htbp]
\begin{center}
\includegraphics[width=60mm]{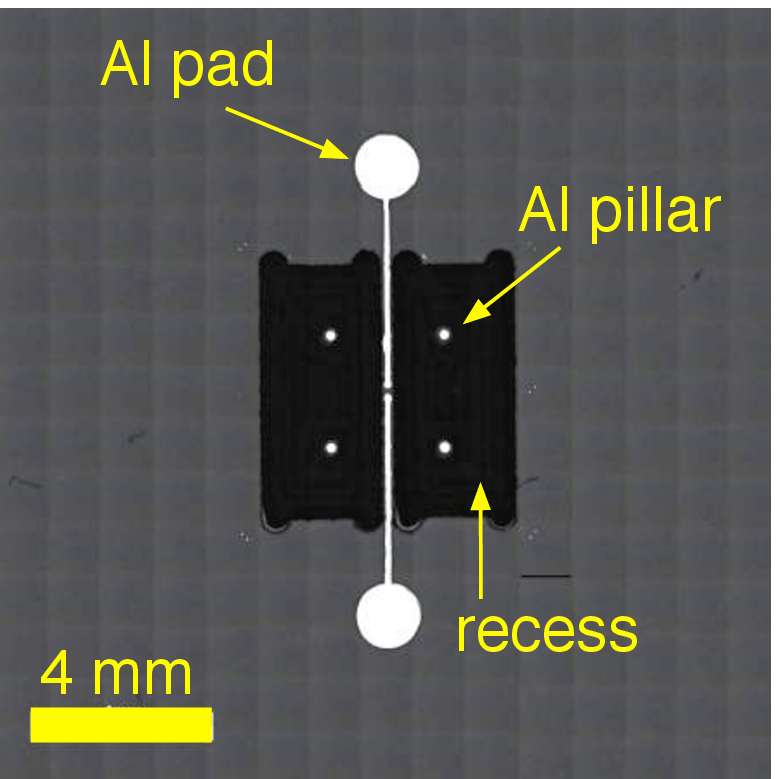} 
\caption{Silica substrate with Al pads, pillars, and a recess.} \label{fig:substrate}
\end{center}
\end{figure}

Counter electrodes made of Al were patterned on a 0.4~mm-thick silica substrate by photo-lithography and wet-etching, as depicted in Fig.~\ref{fig:substrate}. To support the membrane frame with a designed gap of 800 nm between the capacitor electrodes, Al pillars were made on the substrate. In addition, we mechanically carved a burr-free recess with a depth of 50~$\mu$m serving as the pockets of dusts, which can otherwise get stuck between the membrane and the substrate and render the gap between them far bigger than designed. Even with this precaution the actual gap was estimated to be larger ($\approx 1.4~\mu$m) than what was designed (see below).

\section{Optical-cavity design} \label{sec:cavity}
\subsection{A short summary of ray optics}
A Gaussian beam (propagating along the $z$ axis) is characterized completely by a pair of real parameters $z$ and $z_{0}$, or equivalently, a single complex parameter $q$, often called the \textit{q parameter}, defined as
\begin{equation}
  q(z) \equiv z + i z_{0}.
  \label{eq:qParam}
\end{equation}
Here, $z$ represents the distance from the beam waist, i.e., from the position at which the radius of the beam is the minimum.
By convention, $z$ is taken to be positive if the beam waist is on the left side.
$z_{0}$ is called the Rayleigh range, which measures the distance from the beam waist to the position where the beam diameter is $\sqrt{2}$ times larger than that at the beam waist.

Alternatively, the Gauss beam is characterized by an inverse $1/q(z)$,
\begin{equation}
  \frac{1}{q(z)} = \frac{1}{z + i z_{0}} = \frac{1}{R(z)} - i \frac{\lambda}{\pi W^2(z)}
\end{equation}
with
\begin{equation}
  R(z) = z \left[ 1 + \left( \frac{z_{0}}{z} \right)^2 \right]
  \label{eq:Rz}
\end{equation}
being the radius of the curvature of the wavefront, and
\begin{equation}
  W(z) = W_{0} \left[ 1 + \left( \frac{z}{z_{0}} \right)^2 \right]^{1/2}
\end{equation}
being the beam radius, where
\begin{equation}
  W_{0} = \sqrt{ \frac{\lambda z_{0}}{\pi} }
  \label{eq:w0}
\end{equation}
is the waist radius.

\subsection{Optical cavity for EMO NMR}
\begin{figure}[htbp]
\begin{center}
  \includegraphics[width=0.8\linewidth]{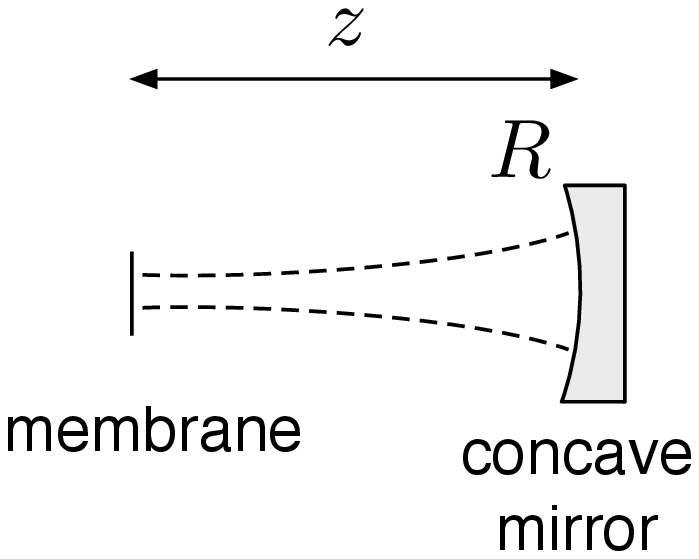}
  \caption{Schematic drawing of an optical cavity.}\label{fig:cavityGeometry}
\end{center}
\end{figure}

An optical cavity, one of key components in the EMO NMR system, is composed of a pair of mirrors, one of which is a metal layer deposited on the membrane, and the other is a concave mirror.
To develop a stable hemispherical laser resonator (Fig.~\ref{fig:cavityGeometry}), the mirrors need to be placed in such a way that the laser beam, which reflects back and forth, form a waist at the membrane, and the wavefront's radius of curvature matches with that of the cavity mirror. In addition, on the concave side of the cavity mirror, the reflection coefficient matches with that (97\%) of the Au mirror on the membrane to achieve the \textit{critical} coupling. Note that due to the other loss mechanism (such as the diffraction loss stemming from the beam misalignment) than the absorption loss of the Au mirror the desired critical coupling could not be realized in the experiment.

In this work, the wavelength $\lambda$ was 780 nm, and we employed the concave mirror with a radius of curvature $R$ of 75 mm, and aimed at setting the beam diameter $2W_{0}$ at the membrane mirror to be 180 $\mu$m so as to safely fit the small Au mirror with the diameter of 0.45~mm. We found, from Eqs.~(\ref{eq:Rz})-(\ref{eq:w0}) that the cavity length $z$ of 17.5 mm fuflils the requirements for the hemispherical resonator.

\section{NMR Experiments} \label{sec:NMR}
In this work, we detected $^{1}$H NMR signals in 0.1 mol/L aqueous solution of CuSO$_4$ containing $\approx 2\times 10^{20}$ $^{1}$H spins. The spin echo experiments were performed by applying successive $\pi/2$ and $\pi$ pulses with a common power of $+17$~dBm, widths of 140 $\mu$s and 280 $\mu$s, and an interval of 1.5 ms.

For generating rf signals and detecting NMR signals, we used a home-built NMR spectrometer equipped with multi-channel rf transmitters and a receiver~\cite{Takeda2008}. Each transmitter is capable of generating rf signals of up to 600 MHz with arbitrary amplitude, phase, frequency, and pulse modulation, and the receiver serves for frequency conversion, digital quadrature demodulation, and digital filtration.

In the conventional electrical detection of the $^{1}$H spin echo demonstrated in the inset of Fig.~\ref{fig:omesignal} for comparison, the rf pulses were fed to port B of Fig.~\ref{fig:setup} while the nuclear induction was detected by monitoring the signal coming out of port A through a low noise amplifier with a noise figure of 1.1 dB. Conversely, in the EMO approach, the drive signal was now fed through port A during acquisition of the photo-detected signal, and the signal from a photo-detector was amplified by another low noise amplifier (SR560, Stanford).
The drive signal was generated with a home-built direct digital synthesizer (DDS) board equipped with a DDS chip AD9858 (Analog Devices)\cite{Takeda2008}.

\begin{figure} [htbp]
\begin{center}
\includegraphics[width=\linewidth]{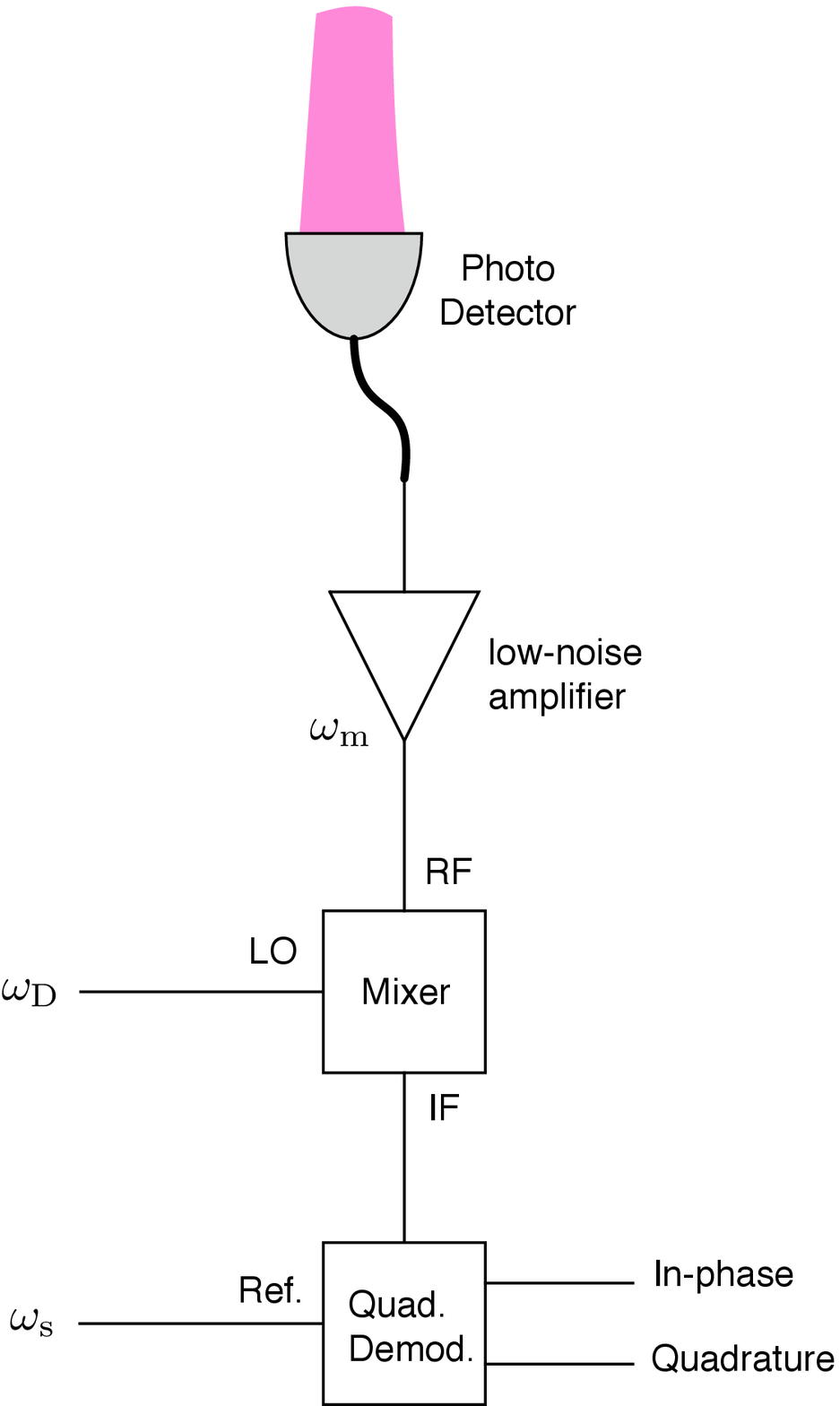} 
\caption{Schematic wiring diagram for quadrature demodulation of the photo-detected nuclear induction signals. The modulation signal at $\omega_{\mathrm{m}}$ from the photo-detector is firstly up-converted with a mixer by the drive at $\omega_{\mathrm{D}}$, and then quadrature demodulated with respect to the rf signal at the NMR frequency $\omega_{\mathrm{s}} = \omega_{\mathrm{D}} - \omega_{\mathrm{m}}$.} \label{fig:demodulation}
\end{center}
\end{figure}

Fig.~\ref{fig:demodulation} shows a schematic diagram of quadrature demodulation of the photo-detected NMR signal with respect to the mechanical frequency $\approx \omega_{\mathrm{m}}$ in such a way that the phase coherence with the excitation rf pulses is retained. Note that the degradation of the signal-to-noise ratio in the process of mandatory signal amplification and frequency conversion is dominated by the amplifier at the first stage, i.e., in the present case, the shot-noise limited photo-detector composed of a Si photo-diode and low-noise operational amplifiers.

\section{Theoretical description}\label{sec:Theory}
To understand the up-conversion mechanism of the signals from rf to optical regimes with the membrane, we recapitulate the theory developed in Ref.~\cite{Polzik2014} with modifications necessary for the present experiments. The theory also serves for evaluation of the signal-to-noise ratio.

\subsection{Hamiltonian}
The Hamiltonian of an LC circuit and a metal-coated membrane oscillator, which is coupled to the former as being a part of the capacitor, can be written as
\begin{equation}
\mathcal{H}(Q,\Phi,Z,P) = \frac{Q^{2}}{2C(Z)} + \frac{\Phi^{2}}{2L} + \frac{m \omega_{0}^{2} Z^{2}}{2}+ \frac{P^{2}}{2 m} - Q V.  \label{eq:U0}
\end{equation}
Here, $Q$, $\Phi$, $C(Z)$, and $L$ are the charge, the flux, the capacitance, and the inductance in the LC circuit. $Z$ and $P$ are the displacement and the momentum of the membrane oscillator, $m$ is the effective mass, and $\omega_{0}/2\pi$ is the eigenfrequency of the \textit{unloaded} membrane oscillator. The coupling between the LC circuit and the membrane oscillator stems from the first term in Eq.~(\ref{eq:U0}) containing the displacement-dependent capacitance. The last term represents the drive applied to the LC circuit with voltage $V$ .

The equations of motion are then
\begin{eqnarray}
\dot{Q} &=& \frac{\partial \mathcal{H}}{\partial \Phi} = \frac{\Phi}{L}\\
\dot{\Phi} &=& -\frac{\partial \mathcal{H}}{\partial Q} = - \frac{Q}{C(Z)} + V \\
\dot{Z} &=& \frac{\partial \mathcal{H}}{\partial P} = \frac{P}{m} \\
\dot{P} &=& -\frac{\partial \mathcal{H}}{\partial Z} = - m \omega_{0}^{2} Z - \frac{Q^{2}}{2} \left( \frac{\partial}{\partial Z} \frac{1}{C(Z)}\right).
\end{eqnarray}
For convenience, let us introduce the following dimensionless variables: $\tilde{Q} =\sqrt[4]{\frac{L}{\hbar^{2} C_{0}}} Q$, $\tilde{\Phi} = \sqrt[4]{\frac{C_{0}}{\hbar^{2} L}} \Phi$, $\tilde{Z} = \sqrt{\frac{m \omega_{0}}{\hbar}} Z$, and $\tilde{P} = \frac{1}{\sqrt{\hbar m \omega_{0}}} P$, where $C_{0}$ is the capacitance with the equilibrium displacement. In addition, we define $C(\tilde{Z}) = \sqrt{\frac{\hbar}{m \omega_{0}}} C(Z) \ [\mathrm{F \cdot m}]$ and $\tilde{V} = \sqrt[4]{\frac{C_{0}}{\hbar^{2} L}} V \ [\mathrm{s}^{-1}]$.
Using the LC resonance frequency $\omega_{\mathrm{LC}}=1/\sqrt{L C_{0}}$, we represent the equations of motion for these variables as
\begin{eqnarray}
\dot{\tilde{Q}} &=& \omega_{\mathrm{LC}} \tilde{\Phi} \label{eq:dQ} \\
\dot{\tilde{\Phi}} &=& - \omega_{\mathrm{LC}} \sqrt{\frac{\hbar}{m \omega_{0}}}  \frac{C_{0}}{C(\tilde{Z})} \tilde{Q} + \tilde{V} \label{eq:dPhi} \\
\dot{\tilde{Z}} &=& \omega_{0} \tilde{P} \label{eq:dX} \\
\dot{\tilde{P}} &=& - \omega_{0} \tilde{Z} - \frac{\tilde{Q}^{2}}{2} \omega_{\mathrm{LC}} \sqrt{\frac{\hbar}{m \omega_{0}}} C_{0} \left( \frac{\partial }{\partial \tilde{Z}} \frac{1}{C(\tilde{Z})}\right). \label{eq:dP}
\end{eqnarray}

Let us now introduce the oscillating drive voltage at $\omega_{\mathrm{D}}$ with the amplitude $V_{0}$, that is, $\tilde{V} = V_{0} \cos \left( \omega_{\mathrm{D}}t \right)$, which poses the major difference from the previous analysis in Ref.~\cite{Polzik2014}, where the DC voltage was applied for realizing the electro-mechanical coupling. To linearize the equations of motion, let us suppose that the mean values and the fluctuations can be separated as $\tilde{Q} = Q_{0} + q$, $\tilde{\Phi} = \Phi_{0} + \phi$, $\tilde{Z} = Z_{0} + z$, and $\tilde{P} = P_{0} + p$. With the ansatz, $Q_{0}=A \cos \left( \omega_{\mathrm{D}}t \right) $ and $\Phi_{0}=B \cos \left( \omega_{\mathrm{D}}t \right)$, we have, from Eqs.~(\ref{eq:dQ}) and (\ref{eq:dPhi}), $A=\left( \omega_{\mathrm{LC}}- \omega_{\mathrm{D}}^{2} / \omega_{\mathrm{LC}} \right)^{-1} V_{0}$, so that
\begin{equation}
Q_{0} = \frac{\omega_{\mathrm{LC}}}{\omega_{\mathrm{LC}}^{2}-\omega_{\mathrm{D}}^{2}} V_{0} \cos \left( \omega_{\mathrm{D}}t \right), \label{eq:Q0}
\end{equation}
which diverges at $\omega_{\mathrm{D}}=\omega_{\mathrm{LC}}$.
The singularity can be avoided if we take dissipation into account. From Eq.~(\ref{eq:dP}), the mean value for the displacement $Z_{0}$ is given by
\begin{align}
Z_{0} & = - \frac{1}{2} \frac{\omega_{\mathrm{LC}}}{\omega_{0}} \sqrt{\frac{\hbar}{m \omega_{0}}} C_{0} \left( \frac{\partial}{\partial \tilde{Z}} \frac{1}{C(Z_{0})} \right) Q_{0}^{2} \nonumber \\
    & = - \frac{1}{2} \frac{\omega_{\mathrm{LC}}}{\omega_{0}} C(Z_{0}) \left( \frac{\partial}{\partial \tilde{Z}} \frac{1}{C(Z_{0})} \right) Q_{0}^{2}
\end{align}
where we use $\sqrt{\frac{\hbar}{m \omega_{0}}} C_{0} = C(Z_{0})$ in the second equation.

For the fluctuations, $q$, $\phi$, $z$, and $p$, the linearized equations of motion around the mean value of the charge $Q_{0}$ and the displacement $Z_{0}$ are given by
\begin{eqnarray}
\dot{q} &=& \omega_{\mathrm{LC}} \phi \label{eq:dQ2} \\
\dot{\phi} &=& - \omega_{\mathrm{LC}}q - \omega_{\mathrm{LC}} \underbrace{C(Z_{0}) \left( \frac{\partial}{\partial z} \frac{1}{C(Z_{0})} \right)}_{\zeta^{-1}} Q_{0} z \label{eq:dPhi2} \\
\dot{z} &=& \omega_{0} p \label{eq:dX2} \\
\dot{p} &=& - \left( \omega_{0} + \underbrace{\frac{1}{2} \omega_{\mathrm{LC}} C(Z_{0}) \left( \frac{\partial^{2}}{\partial z^{2}} \frac{1}{C(Z_{0})} \right) Q_{0}^{2}}_{\delta \omega} \right) z \nonumber \\
       & & - \omega_{\mathrm{LC}} \underbrace{C(Z_{0}) \left( \frac{\partial}{\partial z} \frac{1}{C(Z_{0})} \right)}_{\zeta^{-1}} Q_{0} q. \label{eq:dP2}
\end{eqnarray}
With $Q_{0}$ in Eq.~(\ref{eq:Q0}) and introducing the membrane eigenfrequency $\omega_{\mathrm{m}} \equiv \omega_{0} + \delta \omega$, we have the following linearized equations of motion,
\begin{eqnarray}
\dot{q} &=& \omega_{\mathrm{LC}} \phi \label{eq:dQ3} \\
\dot{\phi} &=& - \omega_{\mathrm{LC}}q - \frac{v_{0}}{\zeta} \cos \left( \omega_{\mathrm{D}}t \right) z \label{eq:dPhi3} \\
\dot{z} &=& \omega_{0} p \label{eq:dX3} \\
\dot{p} &=& - \left( \omega_{\mathrm{m}} \right)  z - \frac{v_{0}}{\zeta} \cos \left( \omega_{\mathrm{D}}t \right) q, \label{eq:dP3}
\end{eqnarray}
where
\begin{equation}
v_{0} = \frac{\omega_{\mathrm{LC}}^{2}}{\omega_{\mathrm{LC}}^{2}-\omega_{\mathrm{D}}^{2}} V_{0}.
\end{equation}

\subsubsection{Linearized Hamiltonian}
The Hamiltonian of the linearized system can be reverse-engineered from Eqs.~(\ref{eq:dQ3}), (\ref{eq:dPhi3}), (\ref{eq:dX3}), and (\ref{eq:dP3}) as
\begin{equation}
H_{0} = \frac{\omega_{\mathrm{LC}}}{2} \left( q^{2} + \phi^{2} \right) + \frac{\omega_{\mathrm{m}}}{2} \left( z^{2} + p^{2} \right) + \frac{v_{0}}{\zeta} \cos \left( \omega_{\mathrm{D}}t \right) q z, \label{eq:H0}
\end{equation}
where we used the fact that $\omega_{\mathrm{m}} \equiv \omega_{0}+\delta \omega \approx \omega_{0}$. By introducing annihilation and creation operators given by $a = (q+i \phi) / \sqrt{2}$, $a^{\dagger} = (q-i \phi) / \sqrt{2}$, $b = (z+i p) / \sqrt{2} $, and $b^{\dagger} = (z-i p) / \sqrt{2}$, we rewrite the Hamiltonian Eq.~(\ref{eq:H0}) as
\begin{equation}
H_{1} = \omega_{\mathrm{LC}} a^{\dagger} a + \omega_{\mathrm{m}} b^{\dagger}b + \frac{v_{0}}{4 \zeta} \left( e^{i \omega_{\mathrm{D}}t} +e^{-i \omega_{\mathrm{D}}t} \right)  \left( a+a^{\dagger} \right) \left( b+b^{\dagger}\right). \label{eq:H1}
\end{equation}
Now by invoking the \textit{rotating-wave approximation} and eliminating the counter-rotating terms $ae^{-i \omega_{\mathrm{D}}}$ and $a^{\dagger}e^{i \omega_{\mathrm{D}}}$, we have
\begin{equation}
H_{2} = \omega_{\mathrm{LC}} a^{\dagger} a + \omega_{\mathrm{m}} b^{\dagger}b + \frac{v_{0}}{4 \zeta} \left( ae^{i \omega_{\mathrm{D}}t} +a^{\dagger}e^{-i \omega_{\mathrm{D}}t} \right) \left( b+b^{\dagger}\right). \label{eq:H2}
\end{equation}
Then, by performing the unitary transformation $ U = \exp \left( i \omega_{\mathrm{D}}t a^{\dagger} a \right) $, we obtain the following time-independent Hamiltonian:
\begin{align}
H_{3} &= UH_{2}U^{\dagger} + i \dot{U} U^{\dagger} \nonumber \\
      & = \left( \omega_{\mathrm{LC}} -\omega_{\mathrm{D}} \right) a^{\dagger} a + \omega_{\mathrm{m}} b^{\dagger}b + \frac{v_{0}}{4 \zeta} \left( a +a^{\dagger} \right) \left( b+b^{\dagger}\right). \label{eq:H3}
\end{align}
This Hamiltonian is recast into the one with the quadratures,
\begin{equation}
H_{\mathrm{em}} = - \frac{\Delta_{\mathrm{i}}}{2} \left( q^{2} + \phi^{2} \right) + \frac{\omega_{\mathrm{m}}}{2} \left( z^{2} + p^{2}  \right) + G_{\mathrm{em}} q z, \label{eq:H4}
\end{equation}
where $-\Delta_{\mathrm{i}}= \omega_{\mathrm{LC}}-\omega_{\mathrm{D}}$ and $G_{\mathrm{em}}$, which describes the electro-mechanical coupling rate, is given by
\begin{equation}
G_{\mathrm{em}}=\frac{v_{0}}{2 \zeta} = \frac{V_{0}}{2} \frac{\omega_{\mathrm{LC}}^{2}}{\omega_{\mathrm{LC}}^{2}-\omega_{\mathrm{D}}^{2}} C(Z_{0}) \left( \frac{\partial}{\partial z} \frac{1}{C(Z_{0})} \right). \label{eq:Gem0}
\end{equation}

By similar arguments, we can establish the effective Hamiltonian for the opto-mechanically coupled system,
\begin{equation}
H_{\mathrm{om}} = -\frac{\Delta_{\mathrm{o}}}{2} \left( X^{2} + Y^{2} \right) + \frac{\omega_{\mathrm{m}}}{2} \left( z^{2} + p^{2}  \right) + G_{\mathrm{om}} X z,
\end{equation}
where $-\Delta_{\mathrm{o}}= \Omega_{\mathrm{c}}-\Omega_{\mathrm{D}}$ with $\Omega_{\mathrm{c}}/2 \pi$ and $\Omega_{\mathrm{D}}/2 \pi$ are the optical cavity frequency and the optical drive frequency, respectively. Here $X$ and $Y$ are the mutually orthogonal quadratures of the intra-cavity optical field, and $G_{\mathrm{om}}$ is the opto-mechanical coupling rate. Putting the electro-mechanically and opto-mechanically coupled systems together, we have the following effective Hamiltonian
\begin{align}
H = & - \frac{\Delta_{\mathrm{i}}}{2} \left( q^{2} + \phi^{2} \right) -\frac{\Delta_{\mathrm{o}}}{2} \left( X^{2} + Y^{2} \right) + \frac{\omega_{\mathrm{m}}}{2} \left( z^{2} + p^{2}  \right) \nonumber \\
    & + G_{\mathrm{em}} q z + G_{\mathrm{om}} X z. \label{eq:H}
\end{align}
Using the Hamiltonian~(\ref{eq:H}), the Heisenberg-Langevin equations of motion, Eqs.~(\ref{eq:q})-(\ref{eq:Y}), are derived with the \textit{ad hoc} dissipation and fluctuation input terms added.


\subsection{Electro-mechano-optical signal transduction}
\subsubsection{rf signal input, Johnson noise, Brownian noise, and optical back-action noise}
The mechanical response, Eq.~(\ref{eq:xF}), contains the emf signal $S$ along with various noise, which can now be rewritten for $\Delta_{\mathrm{i}}=0, \Delta_{\mathrm{o}}=\kappa_{\mathrm{oT}}/2$, and $\omega \rightarrow 0$ using Eqs.~(\ref{eq:q0}) and (\ref{eq:X0}) as
\begin{widetext}
\begin{align}
z(\omega) & = \chi_{\mathrm{m}}(\omega)  \left[ -\sqrt{2 \gamma_{\mathrm{m}}} \underbrace{f_{\mathrm{in}}}_{\mathrm{Brownian\ noise}} \right.
  \left.
  -\frac{G_{\mathrm{em}}}{i\omega -\frac{\kappa_{\mathrm{iT}}}{2}} \left(\sqrt{\gamma_{\mathrm{i}}} \underbrace{q_{\mathrm{in}}}_{\mathrm{Johnson\ noise}} + \sqrt{\kappa_{\mathrm{i}}} \underbrace{Q_{\mathrm{in}}}_{\mathrm{Johnson\ noise}} + \sqrt{\kappa_{\mathrm{i}}} \underbrace{S}_{\mathrm{signal}} \right) \right. \nonumber \\
  & \left. -\frac{G_{\mathrm{om}} \sqrt{\kappa_{\mathrm{o}}}}{\kappa_{\mathrm{oT}}} \left( \underbrace{Y_{\mathrm{in}} -X_{\mathrm{in}}}_{\mathrm{back-action\ noise}} \right) \right]. \label{eq:xT}
\end{align}
\end{widetext}
Here, we see that the rf signal and the Johnson noise are faithfully transduced to the mechanical response with the amplification factor proportional to $G_{\mathrm{em}}$. The added noises are the Brownian noise from the mechnical oscillator and the optical back-action noise, the latter of which can be neglected here since $G_{\mathrm{om}}$ is small. The Brownian noise part cooresponds to the amplifier noise in the convensional NMR detection scheme. The contribution of the Brownian noise can, however, be made negligibly small in principle if the electro-mechanical coupling $G_{\mathrm{em}}$ becomes large. We shall discuss the issue of signal-to-noise in more detail later on in Sec.~\ref{sec:SNR}.

\subsubsection{Optical shot noise and optical signal output}

Now let us see how the mechanical response Eq.~(\ref{eq:xT}) appears in the optical readout. With Eqs.~(\ref{eq:X0}) and (\ref{eq:xT}) with $\Delta_{\mathrm{o}} = \kappa_{\mathrm{oT}}/2$ and $\omega=0$ (this assumption is valid since $\omega \approx \omega_{\mathrm{m}} \ll \Delta_{\mathrm{o}} \ll \Omega_{\mathrm{D}} \approx \Omega_{\mathrm{c}}$), $X$ can be written as
\begin{align}
X =& \frac{\sqrt{\kappa_{\mathrm{o}}}}{\kappa_{\mathrm{oT}}} \left( Y_{\mathrm{in}}-X_{\mathrm{in}} \right) + \frac{G_{\mathrm{om}}}{\kappa_{\mathrm{oT}}} z \nonumber \\
=& \frac{\sqrt{\kappa_{\mathrm{o}}}}{\kappa_{\mathrm{oT}}} \left( Y_{\mathrm{in}}-X_{\mathrm{in}} \right) + \frac{G_{\mathrm{om}}}{\kappa_{\mathrm{oT}}} \chi_{\mathrm{m}}(\omega) \times \nonumber \\
 & \left( -\sqrt{2 \gamma_{\mathrm{m}}}f_{\mathrm{in}} - \frac{G_{\mathrm{em}}}{i \omega -\frac{\kappa_iT}{2}} \left( \sqrt{\kappa_{\mathrm{i}}} \left( Q_{\mathrm{in}} + S \right) + \sqrt{\gamma_{\mathrm{i}}} q_{\mathrm{in}} \right) \right. \nonumber \\
  & \left. -\frac{G_{\mathrm{om}}\sqrt{\kappa_{\mathrm{o}}}}{\kappa_{\mathrm{oT}}} \left(Y_{\mathrm{in}}-X_{\mathrm{in}} \right) \right). \label{eq:optX}
\end{align}
This $X$ has been \textit{tacitly} displaced by $\frac{\alpha + \alpha^{*}}{\sqrt{2}} = {\sqrt{\mathcal{N}_{\mathrm{D}}}}\cos \theta$ from the \textit{lab frame} in the linearized \textit{rotating-frame} Hamiltonian~(\ref{eq:H}), where we only dealt with the fluctuations above the non-zero mean value. Here, $\mathcal{N}_{\mathrm{D}}$ is the intracavity photon number. Since we are in the rotating-frame at $\omega_{\mathrm{D}}$, $\alpha = \sqrt{\frac{\mathcal{N}_{\mathrm{D}}}{2}} e^{i\theta}$ is time-independent.

\section{Signal-to-noise ratio} \label{sec:SNR}
In the conventional electrical detection, the NMR signal acquisition process is inevitably accompanied with amplifier noise, which is characterized by the equivalent amplifier noise temperature $T_{\mathrm{n}} = \sqrt{S_{VV} S_{II}}/k_{\mathrm{B}}$, where $S_{VV}$ and $S_{II}$ are voltage and current noise spectral densities of the amplifier~\cite{HB1997}. For the low-noise amplifier that we used in the conventional electrical NMR detection, $T_{\mathrm{n}}$ was measured to be 84~K. In the EMO approach, where the signals are acquired through membrane displacement measurement, the additional noises come from the Brownian motion of the membrane, the shot noise of the laser beam, and the back-action of the photons hitting on the membrane. Since the last noise is negligible here, the optical shot noise with spectral density $S_{XX}$ and the displacement noise with spectral density $S_{FF}$ contribute to the net noise.

\subsection{Noise spectral densities}

\subsubsection{Brownian noise and Johnson noise}
First, let us evaluate the Johnson noise and the Brownian noise of the mechanical oscillator. From Eq.~(\ref{eq:xT}), the \textit{double-sided} velocity noise spectral density for the mechanical oscillator $S_{\dot{z}\dot{z}}$ is given by
\begin{align}
S_{\dot{z}\dot{z}}(\omega) &= 2 \gamma_{\mathrm{m}} S_{FF}(\omega) + \frac{G_{\mathrm{em}}^{2} \kappa_{\mathrm{iT}}}{\omega^{2} + \frac{\kappa_{\mathrm{iT}}^{2}}{4}} S_{qq}(\omega) \nonumber \\
&= 2 \gamma_{\mathrm{m}} S_{FF}(\omega) + \gamma_{\mathrm{m}} C_{\mathrm{em}}(\omega) S_{qq}(\omega).
\end{align}

The rms displacement noise $\langle z_{\mathrm{n}}^{2} \rangle$ is then given by
\begin{align}
\langle z_{\mathrm{n}}^{2} \rangle \equiv& \int_{0}^{\infty} \frac{d \omega}{2 \pi} \left| \chi_{\mathrm{m}} (\omega) \right|^{2} \left[ S_{\dot{z}\dot{z}}(\omega) + S_{\dot{z}\dot{z}}(-\omega)  \right] \nonumber \\
\approx& \underbrace{ \left( \int_{0}^{\infty} \frac{d \omega}{2 \pi} \frac{\omega_{\mathrm{m}}^{2}}{\left( \omega_{\mathrm{m}}^{2} -\omega^{2} \right)^{2} + \omega^{2}\gamma_{\mathrm{m}}^{2} } \right)}_{\frac{1}{4\gamma_{\mathrm{m}}}} \times \nonumber \\
  & 2 \left( 2 \gamma_{\mathrm{m}} S_{FF}(\omega) + \gamma C_{\mathrm{em}}(\omega) S_{qq}(\omega) \right) \nonumber \\
\approx& \frac{1}{4\gamma_{\mathrm{m}}} \left[ 4 \gamma_{\mathrm{m}} n_{\mathrm{th}}(\omega_{\mathrm{m}},T) + 2 \gamma_{\mathrm{m}} C_{\mathrm{em}}(\omega_{\mathrm{m}}) n_{\mathrm{th}}(\omega_{\mathrm{LC}},T) \right] \nonumber \\
=& n_{\mathrm{th}}(\omega_{\mathrm{m}},T_{\mathrm{eff}}) + \frac{1}{2} C_{\mathrm{em}}(\omega_{\mathrm{m}}) n_{\mathrm{th}}( \omega_{\mathrm{LC}},T). \label{eq:Xn}
\end{align}
We can see, from Eqs.~(\ref{eq:ntm}), (\ref{eq:ntLC}), and (\ref{eq:Xn}), the noise can be dominated by the Johnson noise when
\begin{equation}
\frac{C_{\mathrm{em}}(\omega_{\mathrm{m}})}{2} \left( \frac{\omega_{\mathrm{m}}}{\omega_{\mathrm{LC}}} \right) \left( \frac{T}{T_{\mathrm{eff}}} \right) > 1.
\end{equation}

\subsubsection{Optical shot-noise backaction}
Next, we shall evaluate the optical shot noise. Equation~(\ref{eq:xT}) leads us to the following portion of \textit{double-sided} velocity noise spectral density for the mechanical oscillator, which is from the optical shot noise $S_{XX}(\omega)$ and $S_{YY}(\omega)$:
\begin{equation}
S_{\dot{z}\dot{z}}(\omega) = \frac{G_{\mathrm{om}}^{2}\kappa_{\mathrm{o}}}{\kappa_{\mathrm{oT}}^{2}} (S_{YY}(\omega) +S_{XX}(\omega)) . \label{eq:HX2}
\end{equation}
If these noise spectra $S_{XX}(\omega)$ and $S_{YY}(\omega)$ do not include classical noise around the mechanical frequency $\omega \approx \omega_{\mathrm{m}}$, $S_{XX}(\omega)$ and $S_{YY}(\omega)$ are said to be shot-noise-limited, and
\begin{align}
S_{XX}(\omega) &= n_{\mathrm{th}}(\Omega_{\mathrm{c}},T) + \frac{1}{2}  = \frac{1}{2}, \label{eq:ntSx} \\
S_{YY}(\omega) &= n_{\mathrm{th}}(\Omega_{\mathrm{c}},T) +\frac{1}{2} = \frac{1}{2}, \label{eq:ntSy}
\end{align}
where, since $k_{\mathrm{B}}T \ll \hbar \Omega_{\mathrm{c}}$, $n_{\mathrm{th}}(\Omega_{\mathrm{c}},T)\approx 0$.
Then the noise spectrum of the mechanical displacement is
\begin{widetext}
\begin{align}
\langle z_{\mathrm{s}}^{2} \rangle &= \int_{0}^{\infty} \frac{d \omega}{2 \pi} \left| \chi_{\mathrm{m}} (\omega) \right|^{2} \left[ S_{\dot{z}\dot{z}}(\omega) + S_{\dot{z}\dot{z}}(-\omega) \right] \nonumber \\
&\approx
\underbrace{
  \int_{0}^{\infty} \frac{d \omega}{2 \pi}  \frac{\omega_{\mathrm{m}}^{2}}{\left( \omega_{\mathrm{m}}^{2} -\omega^{2} \right)^{2} + \omega^{2}\gamma_{\mathrm{m}}^{2} }
}_{\frac{1}{4 \gamma_{\mathrm{m}} } }
\frac{2 G_{\mathrm{om}}^{2}\kappa_{\mathrm{o}}}{\kappa_{\mathrm{oT}}^{2}} (S_{YY}(\omega) +S_{XX}(\omega)) \nonumber \\
&= \frac{1}{4 \gamma_{\mathrm{m}}} \frac{2 G_{\mathrm{om}}^{2}\kappa_{\mathrm{o}}}{\kappa_{\mathrm{oT}}^{2}} \left( \frac{1}{2} + \frac{1}{2} \right)  \nonumber \\
&= \frac{G_{\mathrm{om}}^{2} \kappa_{\mathrm{o}}}{2 \gamma_{\mathrm{m}} \kappa_{\mathrm{oT}}^{2}} = \frac{C_{\mathrm{om}}}{2} \frac{\kappa_{\mathrm{o}}}{\kappa_{\mathrm{oT}}}, \label{eq:Xs}
\end{align}
where the \textit{opto-mechanical cooperativity} $C_{\mathrm{om}}$ is
\begin{equation}
C_{\mathrm{om}} = \frac{G_{\mathrm{om}}^{2}}{\gamma_{\mathrm{m}} \kappa_{\mathrm{oT}}}. \label{eq:Com}
\end{equation}

The total mechanical displacement driven by various noise sources, the Brownian noise and the Johnson noise [see Eq.~(\ref{eq:Xn})] and the shot noise [see Eq.~(\ref{eq:Xs})] is
\begin{align}
\langle z_{\mathrm{T}}^{2} \rangle =& \langle z_{\mathrm{n}}^{2} \rangle + \langle z_{\mathrm{s}}^{2} \rangle \nonumber \\
=& \int_{0}^{\infty} \frac{d \omega}{2 \pi} \left| \chi_{\mathrm{m}} (\omega) \right|^{2}  \gamma_{\mathrm{m}} \left[ S_{FF} (\omega) + \frac{1}{2} C_{\mathrm{em}}(\omega) S_{qq}(\omega)
     + \frac{1}{2} C_{\mathrm{om}}\frac{\kappa_{\mathrm{o}}}{\kappa_{\mathrm{oT}}} \left( S_{XX}(\omega) + S_{YY}(\omega) \right) \right] \nonumber \\
=& n_{\mathrm{th}}(\omega_{\mathrm{m}},T) + \frac{1}{2} C_{\mathrm{em}}(\omega_{\mathrm{m}})  n_{\mathrm{th}}(\omega_{\mathrm{LC}},T) + \frac{1}{2} C_{\mathrm{om}}\frac{\kappa_{\mathrm{o}}}{\kappa_{\mathrm{oT}}}
\end{align}

\subsection{Signal-to-noise ratio}
In the presence of appreciable phase noise of the drive signal, the single-sided spectral density at frequency $\omega$ in Eq.~(\ref{eq:Soo}) is modified to be
\begin{align}
S_{\mathrm{oo}}(\omega) =& \kappa_{\mathrm{o}} \mathcal{N}_{\mathrm{D}} \left[ \left( \left(\frac{\kappa_{\mathrm{o}}}{\kappa_{\mathrm{oT}}} \right)^{2} +  \left(1-\frac{\kappa_{\mathrm{o}}}{\kappa_{\mathrm{oT}}} \right)^{2} \right) \left( 2S_{XX}(\omega) + 2S_{YY}(\omega) \right) \right. \nonumber \\
& + C_{\mathrm{om}}\frac{\kappa_{\mathrm{o}}}{\kappa_{\mathrm{oT}}} 2 \gamma_{\mathrm{m}}^{2} \left| \chi_{\mathrm{m}}(\omega) \right|^{2} 4S_{FF}(\omega) \nonumber \\
& \left. + C_{\mathrm{om}}\frac{\kappa_{\mathrm{o}}}{\kappa_{\mathrm{oT}}} C_{\mathrm{em}}(\omega) \gamma_{\mathrm{m}}^{2} \left| \chi_{\mathrm{m}}(\omega) \right|^{2} \left[ 4S_{qq}(\omega) + 4 \frac{\kappa_{\mathrm{i}}}{\kappa_{\mathrm{iT}}} S^{2} \delta \left( \omega-\omega_{\mathrm{m}} \right)  + 4 \frac{\kappa_{\mathrm{i}}}{\kappa_{\mathrm{iT}}} \mathcal{L}(\omega) \frac{P_{\mathrm{D}}}{\hbar \omega_{\mathrm{D}}} \right] \right]. \label{eq:Soo2}
\end{align}
By taking the phase noise into account, the signal-to-noise ratio, Eq.~(\ref{eq:SNR}), is modified as
\begin{align}
  \frac{S}{N} =
 \sqrt{\frac{S^{2}}{\cfrac{\kappa_{\mathrm{iT}}}{\kappa_{\mathrm{i}}} \left( \cfrac{S_{XX}(\omega_{\mathrm{m}}) + S_{YY}(\omega_{\mathrm{m}}) }{2C_{\mathrm{om}} \frac{\kappa_{\mathrm{o}}}{\kappa_{\mathrm{oT}}} C_{\mathrm{em}}(\omega_{\mathrm{m}}) } + \cfrac{2S_{FF}(\omega_{\mathrm{m}})}{C_{\mathrm{em}}(\omega_{\mathrm{m}})} + S_{qq}(\omega_{\mathrm{m}}) + \cfrac{\kappa_{\mathrm{i}}}{\kappa_{\mathrm{iT}}}\mathcal{L}(\omega_{\mathrm{m}}) \cfrac{P_{\mathrm{D}}}{\hbar \omega_{\mathrm{D}}} \right) \Delta}}. \label{eq:SNR3}
\end{align}
\end{widetext}

\section{Parameter calibrations} \label{sec:parameters}
\subsection{Equilibrium distance $d_{0}$ between the electrodes of the membrane capacitor} \label{sec:d0}

The electro-mechanical cooperativity $C_{\mathrm{em}}$ can be deduced from the shift of the eigenfrequency of the membrane oscillator as a function of the drive power. The frequency shift $\delta \omega$ is given with the dimensionless variables by
\begin{equation}
\delta \omega = \frac{1}{2} \omega_{\mathrm{LC}} C(Z_{0}) \left( \frac{\partial^{2}}{\partial z^{2}} \frac{1}{C(Z_{0})} \right) Q_{0}^{2}
\end{equation}
as in Eq.~(\ref{eq:dP2}). Here $Q_{0}$ is the one given by Eq.~(\ref{eq:Q0}) with a loss term, that is,
\begin{equation}
Q_{0} = \frac{\omega_{\mathrm{LC}}}{ \left( \omega_{\mathrm{LC}}^{2}-\omega_{\mathrm{D}}^{2} \right) -i \omega_{\mathrm{LC}} \kappa_{\mathrm{iT}}} V_{0} \cos \left( \omega_{\mathrm{D}} t \right).
\end{equation}
At resonance $\Delta_{\mathrm{i}}=\omega_{\mathrm{D}}-\omega_{\mathrm{LC}}=0$,
\begin{equation}
Q_{0} = i \frac{V_{0}}{\kappa_{\mathrm{iT}}} \cos \left( \omega_{\mathrm{D}} t \right). \label{eq:Q0_2}
\end{equation}
On the other hand, for $Q_{0}^{2}$ the dominant contribution comes from the rectified DC term, that is,
\begin{equation}
Q_{0}^{2} = \frac{1}{2} \frac{V_{0}^{2}}{\omega_{\mathrm{LC}}^{2}}. \label{eq:Q0_3}
\end{equation}

\begin{figure} [bt]
\begin{center}
\includegraphics[width=\linewidth]{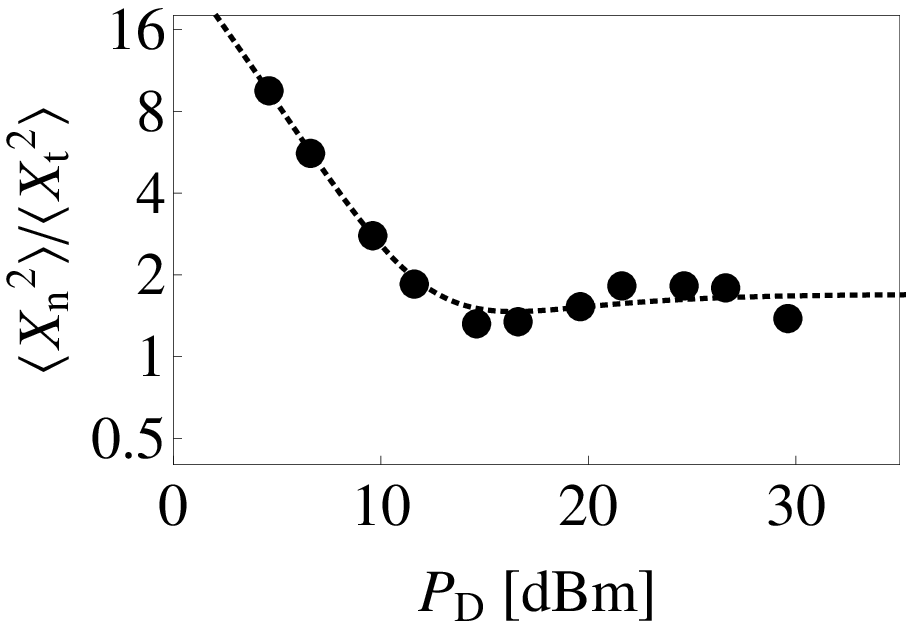}
\caption{Ratio $\langle X_{\mathrm{n}}^{2} \rangle / \langle X_{\mathrm{t}}^{2} \rangle$ as a function of the drive power $P_{\mathrm{D}}$. The points are the experimental values obtained from the data shown in Fig.~3. The dashed line represents the best fit based on Eq.~(\ref{eq:R}) where the fitting parameters are $T_{\mathrm{eff}}$ contained in $n_{\mathrm{th}}(\omega_{\mathrm{m}},T_{\mathrm{eff}})$, $\mathcal{L}(\omega_{\mathrm{m}}+\Delta_{\mathrm{T}})$, and $\eta_{p}$.} \label{fig:cal_r}
\end{center}
\end{figure}

\begin{figure} [htbp]
\begin{center}
\includegraphics[width=\linewidth]{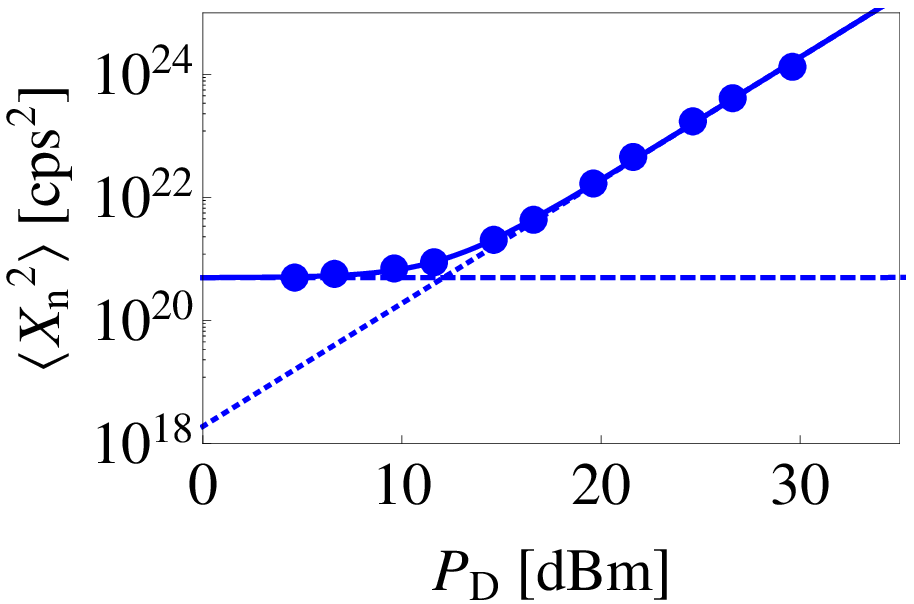} 
\caption{$\langle X_{\mathrm{n}}^{2} \rangle$ as a function of the drive power $P_{\mathrm{D}}$. The points represent the experimentally obtained values of $\langle X_{\mathrm{n}}^{2} \rangle$, which is calibrated in units of (photon number flux)$^2$ based on the shot noise level as a reference. The thick line represents the best fit based on Eq.~(\ref{eq:Sn}), where the fitting parameter is just $C_{\mathrm{om}}$ as the other parameters are \textit{a priori} given. The dashed line represents the contribution from the sum of the Johnson noise and the mechanical Brownian noise, while the dotted line represents the contribution from the phase noise of the drive in Eq.~(\ref{eq:Sn}).} \label{fig:cal_n}
\end{center}
\end{figure}

\begin{figure} [htbp]
\begin{center}
\includegraphics[width=\linewidth]{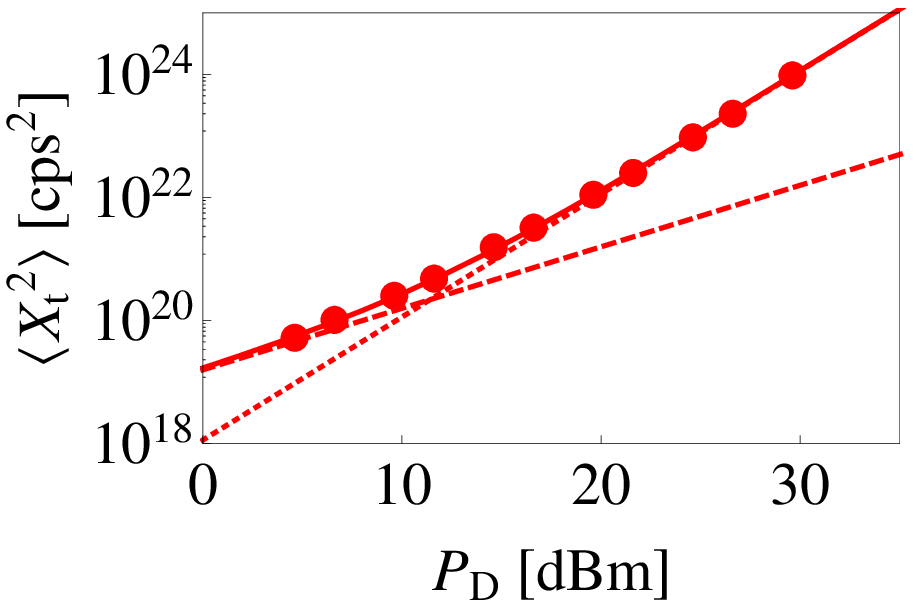} 
\caption{$\langle X_{\mathrm{t}}^{2} \rangle$ as a function of the drive power $P_{\mathrm{D}}$. The points represent the experimentally obtained values of $\langle X_{\mathrm{t}}^{2} \rangle$, which is calibrated in units of (photon number flux)$^2$ based on the shot noise level as a reference. The thick line represents the best fit based on Eq.~(\ref{eq:St}), where the fitting parameter is just $C_{\mathrm{om}}$ as the other parameters are \textit{a priori} given. The dashed line represents the contribution from the tone signal, while the dotted line represents the contribution from the phase noise of the drive in Eq.~(\ref{eq:St}).} \label{fig:cal_t}
\end{center}
\end{figure}

%
With this $Q_{0}^{2}$ the frequency shift can be written with the \textit{dimensionless} variables as
\begin{equation}
\delta \omega \approx -\frac{1}{4} \left(\frac{1}{C(Z_{0})} \frac{\partial^{2} C(Z_{0})}{\partial z^{2}} \frac{V_{0}^{2}}{\omega_{\mathrm{LC}}} \right),
\end{equation}
where we use $\frac{\partial^{2}}{\partial z^{2}} \frac{1}{C(Z_{0})} \approx -\frac{1}{C(Z_{0})^{2}} \frac{\partial^{2} C(Z_{0})}{\partial z^{2}}$, which holds in the case where the contribution of the displacement-dependent membrane capacitance to the total capacitance is small. With the \textit{physical} variables, $\delta \omega$ can be given by
\begin{equation}
\delta \omega \approx -\frac{1}{4 m \omega_{0}} \frac{\partial^{2} C (Z)}{\partial Z^{2}} V^{2}. \label{eq:domega}
\end{equation}
The capacitance $C(Z)$ is the total capacitance of a series LCR circuit that is equivalent to the impedance-matched probe LCR circuit shown in Fig.~1. The total capacitance is the sum, $C(Z)=C_{\mathrm{t}} + C_{\mathrm{p}} + C_{\mathrm{m}}(Z)$, of the trimmer capacitance being $C_{\mathrm{t}} \approx 98$~pF, the parasitic capacitance being $C_{\mathrm{p}} \approx 21$~pF and the membrane capacitance $C_{\mathrm{m}}(Z)$. Here, the membrane capacitor can be approximated as two parallel plate capacitors in series and its capacitance is given by
\begin{equation}
C_{\mathrm{m}}(Z) = \frac{\epsilon_{0}}{2} \frac{A}{d+Z}
\end{equation}
with $\epsilon_{0}$ being the vacuum permittivity, $A$ being the area of the capacitor, and $d$ is the \textit{nominal} distance between the electrodes. Then Eq.~(\ref{eq:domega}) becomes
\begin{equation}
\delta \omega \approx -\frac{R P_{\mathrm{D}}}{4 m \omega_{0}} \frac{\epsilon_{0} A}{d_{0}^{3}}, \label{eq:domega2}
\end{equation}
where $d_{0} =d+Z_{0}$ is the equilibrium distance between the electrodes for a given power $P_{\mathrm{D}}=V^{2}/R$ of the drive and $R=50$~$\Omega$ is the impedance of the circuit looking from port A at the drive frequency.

With the independently estimated parameters, $m=8.6 \times 10^{-11}$~kg, $\omega_{0}/2 \pi = 180$~kHz, $A= \pi \times (160~\mu\mathrm{m}/2)^{2}$, we can deduce $d_{0} \approx 1.4\ \mu\mathrm{m}$ from the drive-power dependence of the membrane resonance frequency (see Fig.~\ref{fig:extone}). This value is almost twice as large as what we designed (800~nm). Presumably, dusts and/or substrate deformations within the membrane capacitor cause the discrepancy.

\subsection{Electro-mechanical cooperativity $C_{\rm{em}}$}
%
%
The electro-mechanical coupling rate $G_{\mathrm{em}}$ can then be estimated. From Eqs.~(\ref{eq:dP2}), (\ref{eq:Gem0}), and (\ref{eq:Q0_2}) we have
\begin{equation}
\frac{G_{\mathrm{em}}}{\delta \omega} = \frac{\frac{v_{0}}{2\zeta}}{\delta \omega} = \frac{ \frac{\partial}{\partial z} \frac{1}{C(Z_{0})}}{\frac{\partial^{2}}{\partial z^{2}}\frac{1}{C(Z_{0})}} \frac{\omega_{\mathrm{LC}}^{2}}{\kappa_{\mathrm{iT}} V_{0}},
\end{equation}
with the dimensionless variables. Thus, with the \textit{physical} variables, the electro-mechanical coupling rate can be given by
\begin{equation}
G_{\mathrm{em}} = \left( \frac{\omega_{\mathrm{LC}}^{2}}{\kappa_{\mathrm{iT}}V} \right) \sqrt{m \omega_{0}} \sqrt[4]{\frac{L}{C_{0}}} \left( \frac{\frac{\partial}{\partial Z} \frac{1}{C(Z)}}{\frac{\partial^{2} }{\partial Z^{2}} \frac{1}{C(Z)}} \right) \delta \omega
\end{equation}
Using the approximate identity~\cite{Polzik2014},
\begin{equation}
\left( \frac{\partial}{\partial Z} \frac{1}{C(Z)} \right) = - \left( \frac{\partial^{2}}{\partial Z^{2}} \frac{1}{C(Z)} \right) \frac{d_{0}}{2},
\end{equation}
and Eq.~(\ref{eq:domega2}) we have
\begin{eqnarray}
G_{\mathrm{em}} &=& \left( \frac{\omega_{\mathrm{LC}}^{2}}{\kappa_{\mathrm{iT}} V} \right) \sqrt{m \omega_{0}} \sqrt[4]{\frac{L}{C_{0}}} \frac{d_{0}}{2} \left(\frac{R P_{\mathrm{D}}}{4 m \omega_{0}} \frac{\epsilon_{0} A}{d_{0}^{3}} \right) \nonumber \\
&=& \frac{1}{2}\omega_{\mathrm{LC}} \frac{z_{\mathrm{zpf}}}{2 d_{0}} \frac{C_{m}(Z)}{C_{0}} \sqrt{\frac{P_{\mathrm{D}}}{\hbar \omega_{\mathrm{LC}} \kappa_{\mathrm{iT}}}} \nonumber \\
&=& \frac{1}{2} g_{\mathrm{em}} \sqrt{\frac{P_{\mathrm{D}}}{\hbar \omega_{\mathrm{LC}} \kappa_{\mathrm{iT}}}}, \label{GemR}
\end{eqnarray}
where we used $2R/L = \kappa_{\mathrm{iT}}$ and $z_{\mathrm{zpf}} = \sqrt{\hbar/(2 m \omega_{0})}$, which is the zero point fluctuation of the membrane oscillator. The form of $G_{\mathrm{em}}$ has a clear  physical explanation; $\frac{1}{2}$ stems from the rotating-wave approximation we have performed in Eq.~(\ref{eq:H2}), $g_{\mathrm{em}} \equiv \omega_{\mathrm{LC}} (z_{\mathrm{zpf}}/(2d_{0})) (C_{\mathrm{m}}(Z)/C_{0}) $ is the so-called \textit{single-photon electro-mechanical coupling rate}~\cite{Clerk2010}, where the multiplication factor $\eta \equiv C_{\mathrm{m}}(Z)/C_{0}$ signifies the contribution of the membrane capacitor to the total capacitance $C_{0}$, and $\sqrt{P_{\mathrm{D}}/(\hbar \omega_{\mathrm{LC}} \kappa_{\mathrm{iT}}})$ is the square root of the intra-LC resonator photon number in the case of resonant drive.

With the following parameters, $d_{0} = 1.4\ \mu\mathrm{m}$, $\omega_{\mathrm{LC}}/2 \pi = 38$~MHz, $z_{\mathrm{zpf}} = 7.3\times 10^{-16}$~m, and $\eta=0.52 \times 10^{-3}$, we have $g_{\mathrm{em}}/2 \pi \approx 5.1 \times 10^{-6}$~Hz and $G_{\mathrm{em}}/2 \pi = 0.9$~kHz for $P_{\mathrm{D}}=+15$~dBm. We can now estimate the electro-mechanical cooperativity $C_{\mathrm{em}}$ from the definition given by Eq.~(\ref{eq:Cem}). With the independently measured values of $\kappa_{\mathrm{iT}}/2 \pi=1.6$~MHz ($\gamma_{\mathrm{i}}/2 \pi = 800$~kHz; $\kappa_{\mathrm{i}}/2 \pi = 810$~kHz) and $\gamma_{\mathrm{m}}/2 \pi= 100$~Hz, we have $C_{\mathrm{em}} (\omega_{\mathrm{m}}) = 0.019$ for $P_{\mathrm{D}}=+15$~dBm.

\subsection{Phase noise and mechanical bath temperature}

Now let us evaluate the phase noise and the mechanical bath temperature. To this end, we examine how the noise and signal grow as a function of the electro-mechanical cooperativity
$C_{\mathrm{em}}(\omega)
\propto P_{\mathrm{D}}$.
Here we neglect the shot noise and any other noise contributing to the noise floor as it can be subtracted from the data. As we have seen in Eq.~(\ref{eq:Soo2}), a sum of the Johnson noise, the mechanical Brownian noise, and the phase noise of the drive lead to the following form
\begin{widetext}
\begin{equation}
S_{\mathrm{n}}(\omega) = \kappa_{\mathrm{o}} \mathcal{N}_{\mathrm{D}} \left[C_{\mathrm{om}}\frac{\kappa_{\mathrm{o}}}{\kappa_{\mathrm{oT}}} 2 \gamma_{\mathrm{m}}^{2} \left| \chi_{\mathrm{m}}(\omega) \right|^{2} 4S_{FF}(\omega) + C_{\mathrm{om}}\frac{\kappa_{\mathrm{o}}}{\kappa_{\mathrm{oT}}} C_{\mathrm{em}}(\omega) \gamma_{\mathrm{m}}^{2} \left| \chi_{\mathrm{m}}(\omega) \right|^{2} \left( 4S_{qq}(\omega) + 4 \frac{\kappa_{\mathrm{i}}}{\kappa_{\mathrm{iT}}}\mathcal{L}(\omega) \frac{P_{\mathrm{D}}}{\hbar \omega_{\mathrm{D}}} \right) \right]. \label{eq:Sn}
\end{equation}
When applying a \textit{narrow band} external tone at $\omega_{\mathrm{T}} = \omega_{\mathrm{D}} + \omega_{\mathrm{m}} + \Delta_{\mathrm{T}}$ with the power $P_{\mathrm{T}}$, on the other hand, the resultant spectrum is given by
\begin{equation}
S_{t}(\omega) =
  \kappa_{\mathrm{o}} \mathcal{N}_{\mathrm{D}} \left[
    C_{\mathrm{om}}\frac{\kappa_{\mathrm{o}}}{\kappa_{\mathrm{oT}}} C_{\mathrm{em}}(\omega) \gamma_{\mathrm{m}}^{2} \left| \chi_{\mathrm{m}}(\omega) \right|^{2} \frac{\kappa_{\mathrm{i}}}{\kappa_{\mathrm{iT}}} \left( 4 \frac{P_{\mathrm{T}}}{\hbar \omega_{\mathrm{T}}} \delta \left( \omega- \omega_{\mathrm{m}} - \Delta_{\mathrm{T}} \right) + 4 \mathcal{L}(\omega) \frac{P_{\mathrm{D}}}{\hbar \omega_{\mathrm{D}}} \right)
  \right]. \label{eq:St}
\end{equation}
Comparing these noise powers (area) we have
\begin{eqnarray}
\frac{\langle X_{\mathrm{n}}^{2} \rangle}{\langle X_{\mathrm{t}}^{2} \rangle} &\equiv& \frac{\int_{\omega_{\mathrm{m}}-\frac{\Delta}{2}}^{\omega_{\mathrm{m}}+\frac{\Delta}{2}} \frac{d \omega}{2 \pi} S_{\mathrm{n}}(\omega) }{\int_{\omega_{\mathrm{m}}+\Delta_{\mathrm{T}}-\frac{\delta}{2}}^{\omega_{\mathrm{m}}+\Delta_{\mathrm{T}}+\frac{\delta}{2}} \frac{d \omega}{2 \pi} S_{\mathrm{t}}(\omega) }\nonumber \\
&=& \frac{ 2 \gamma_{\mathrm{m}} n_{\mathrm{th}}(\omega_{\mathrm{m}},T) + C_{\mathrm{em}}(\omega_{\mathrm{m}}) \left( \gamma_{\mathrm{m}} n_{\mathrm{th}}(\omega_{\mathrm{LC}},T) + \frac{\kappa_{\mathrm{i}}}{\kappa_{\mathrm{iT}}} \eta_{p} \frac{P_{\mathrm{D}}}{\hbar \omega_{\mathrm{D}}} \right)}{\gamma_{\mathrm{m}}^{2} \left| \chi_{\mathrm{m}}(\omega_{\mathrm{m}}+\Delta_{\mathrm{T}}) \right|^{2} C_{\mathrm{em}}(\omega_{\mathrm{m}}+\Delta_{\mathrm{T}}) \frac{\kappa_{\mathrm{i}}}{\kappa_{\mathrm{iT}}} \left( \frac{P_{\mathrm{T}}}{\hbar \omega_{\mathrm{T}}} + \mathcal{L}(\omega_{\mathrm{m}}+\Delta_{\mathrm{T}}) \delta \frac{P_{\mathrm{D}}}{\hbar \omega_{\mathrm{D}}} \right)}, \label{eq:R}
\end{eqnarray}
\end{widetext}
which does not contain the optical quantities, $\mathcal{N}_{\mathrm{D}}$, $C_{\mathrm{om}}$, $\kappa_{\mathrm{o}}$, $\kappa_{\mathrm{oT}}$, where the range of the integration for the denominator $\delta$ is a little bit more than the bandwidth over which the tone signal is appreciable thus the phase noise contribution can be modeled as $\mathcal{L}(\omega_{\mathrm{m}}+\Delta_{\mathrm{T}}) \delta \frac{P_{\mathrm{D}}}{\hbar \omega_{\mathrm{D}}}$, while the range of the integration for the numerator $\Delta$ ($\Delta \gg \delta$) is a little bit more than the bandwidth of the mechanical response $\gamma_{\mathrm{m}}$ thus the phase noise contribution becomes
\begin{equation}
\eta_{p} = \int_{\omega_{\mathrm{m}}-\frac{\Delta}{2}}^{\omega_{\mathrm{m}}+\frac{\Delta}{2}} \frac{d \omega}{2 \pi} \gamma_{\mathrm{m}}^{2} \left| \chi_{\mathrm{m}}(\omega) \right|^{2} \mathcal{L}(\omega).
\end{equation}

Figure~\ref{fig:cal_r} shows the ratio $\langle X_{\mathrm{n}}^{2} \rangle / \langle X_{\mathrm{t}}^{2} \rangle$ as a function of the drive power $P_{\mathrm{D}}$, which is generated from Fig.~3. From this data, the mechanical bath temperature given by Eq.~(\ref{eq:ntm}) is estimated to be $T_{\mathrm{eff}} \approx 205$~K, which is more or less consistent with the environment temperature of 300~K, indicating that there are no appreciable heating effect.  As for the phase noise, we deduce $\mathcal{L}(\omega_{\mathrm{m}}+\Delta_{\mathrm{T}})\delta \approx 5.8 \times 10^{-10}$, and $\eta_{p} \approx 9.6 \times 10^{-12}$ from the data. The former corresponds to the phase noise bandwidth [see Eq.~(\ref{eq:L}) for definition] $\delta_{\mathrm{P}}/2 \pi \approx 19$~Hz, while the latter does to $\delta_{\mathrm{P}}/2 \pi \approx 31$~Hz. These values may be sensible given that our model of the phase noise given in Eq.~(\ref{eq:L}) is a simple one ignoring $1/f$ noise and frequency-independent noise.

\subsection{Opto-mechanical cooperativity $C_{\rm{om}}$}
%
%

\begin{table*}[htbp]
 \caption{Noise budget of the prospective EMO NMR detection.}
  \begin{center}
   \begin{tabular}{|c||c|c|c||c|}
    \hline
      \ & Shot noise & Brownian noise & Johnson noise & Total noise \\
    \hline \hline
     Symbolic notation$^{\cfrac{}{}}_{\cfrac{}{}}$ & \ $\cfrac{S_{XX}+S_{YY}}{2C_{\mathrm{om}} C_{\mathrm{em}}} $\  & $\cfrac{2 S_{FF}}{C_{\mathrm{em}}}$ &  $ S_{qq}$ &  \\
    \hline
      Number of quanta & $ 0.069 $ & $ 3100 $ & $1.6 \times 10^5$ & $ 1.7 \times 10^{5}$ \\
    \hline
      \ Effective temperature [K] \ & $1.3 \times 10^{-4}$ & 5.6  & 300 & 306  \\
    \hline
   \end{tabular}
  \end{center} \label{tb:p}
\end{table*}

The opto-mechanical coupling rate $G_{\mathrm{om}}$ can now be estimated. The strategy is to use the optical shot noise level as a reference~\cite{Hisatomi2016} and evaluate the experimentally obtained noise spectral density $\langle X_{\mathrm{n}}^{2} \rangle$ and that for the tone $\langle X_{\mathrm{t}}^{2} \rangle$ in Eq.~(\ref{eq:R}) from the data shown in Fig.~3. Blue points in Fig.~\ref{fig:cal_n} represent $\langle X_{\mathrm{n}}^{2} \rangle$ calibrated in units of (photon number flux)$^2$ at respective drive power $P_{\mathrm{D}}$. From this data and with already known parameters, we deduce the opto-mechanical cooperativity $C_{\mathrm{om}} \approx 0.32 \times 10^{-3}$. Red points in Fig.~\ref{fig:cal_t}, on the other hand, represent $\langle X_{\mathrm{t}}^{2} \rangle$ calibrated in units of (photon number flux)$^2$ at respective drive power $P_{\mathrm{D}}$. From this data, we deduce $C_{\mathrm{om}} \approx 0.33 \times 10^{-3}$, which is in good agreement with the former value.

We can then estimate $G_{\mathrm{om}}$ from the definition given by Eq.~(\ref{eq:Com}). With the independently measured values of $\kappa_{\mathrm{oT}}/2 \pi=1.1$~GHz  ($\gamma_{\mathrm{o}}/2 \pi = 1.1$~GHz; $\kappa_{\mathrm{o}}/2 \pi = 43$~MHz) and $\gamma_{\mathrm{m}}/2 \pi= 100$~Hz, we obtain $G_{\mathrm{om}}/2\pi = 6.0$~kHz. The \textit{single-photon opto-mechanical coupling rate} $g_{\mathrm{om}}$ is given by
\begin{equation}
G_{\mathrm{om}} = \frac{1}{2} g_{\mathrm{om}} \sqrt{\mathcal{N}_{\mathrm{D}}} = \frac{1}{2} g_{\mathrm{om}} \sqrt{\frac{\mathcal{P}_{\mathrm{D}}}{\hbar \Omega_{\mathrm{D}}} \frac{2 \kappa_{\mathrm{o}}}{\kappa_{\mathrm{oT}}^{2}}}.
\end{equation}
With the above $G_{\mathrm{om}}$ for the optical input power of $\mathcal{P}_{\mathrm{D}}=1.2$~mW, we have $g_{\mathrm{om}}/2 \pi \approx 55$~Hz. When the opto-mechanical coupling purely stems from the radiation pressure, $g_{\mathrm{om}}$ can be given by~\cite{Clerk2010}
\begin{equation}
g_{\mathrm{om}} = \Omega_{\mathrm{c}} \frac{z_{\mathrm{zpf}}}{l}
\end{equation}
with $l \approx 18$~mm being the cavity length, which results in $g_{\mathrm{om}}/2 \pi \approx 16$~Hz. Thus the opto-mechanical coupling could be partly due to the radiation pressure and partly due to the photo-thermal effect~\cite{Karrai2004}.

\section{Prospect} \label{sec:prospect}

There is plenty of room for reducing the added noises with realistic improvements in the parameters. For instance, if the Au layer coated on the membrane is replaced by an aluminum layer, the weight would be reduced by a factor of $\approx 7$, and the frequency of the mechanical resonance would be much higher. Then, the phase noise of the drive at $\omega_{\mathrm{m}}$ can be significantly smaller. Moreover, we could then use a notch filter that prevent the phase noise around $\omega_{\mathrm{m}}$ from entering the LC circuit. The drive power $P_{\mathrm{D}}$ of $+30$~dBm could then be applied for increasing $C_{\mathrm{em}} \propto P_{\mathrm{D}}$.


If the cavity and the LC circuit are assumed to be both overcoupled, i.e., $\kappa_{\mathrm{o}} \approx \kappa_{\mathrm{oT}}$ and $\kappa_{\mathrm{i}} \approx \kappa_{\mathrm{iT}}$, the signal-to-noise ratio Eq.~(\ref{eq:SNR}) can be simplified to
\begin{widetext}
\begin{equation}
\frac{S}{N} = \sqrt{\frac{S^{2}}{ \left( \cfrac{S_{XX}(\omega_{\mathrm{m}})+S_{YY}(\omega_{\mathrm{m}})}{2C_{\mathrm{om}} C_{\mathrm{em}}(\omega_{\mathrm{m}}) } + \cfrac{2 S_{FF}(\omega_{\mathrm{m}})}{C_{\mathrm{em}}(\omega_{\mathrm{m}})} + S_{qq}(\omega_{\mathrm{m}}) \right) \Delta}}. \label{eq:SNR2}
\end{equation}
\end{widetext}
We can see that as the electro-mechanical cooperativity $C_{\mathrm{em}}$ increases, the contributions of the shot noise, $S_{XX}+S_{YY}$, and the mechnical Brownian noise, $S_{FF}$, to the total noise decrease, and the dominant noise would be the intrinsic Johnson noise, $S_{qq}$, from the LC circuit. In addition, the larger opto-mechanical cooperativity $C_{\mathrm{om}}$ is beneficial to minimize the shot noise contribution further.

A major improvement of the signal-to-noise ratio can be achieved by reducing the gap of the membrane capacitor $d_{0}$. Suppose that $d_{0}$ is changed from the current value of $ d_{0} \approx 1.4~\mu$m to 100~nm, the electro-mechanical coupling scales as $G_{\mathrm{em}} \propto 1/d_{0}^{2}$ thereby $C_{\mathrm{em}} \propto 1/d_{0}^{4}$ changes from the current value of $\approx 0.02$ to $\approx 700$. With the drive power of $+30$~dBm, $C_{\mathrm{em}}$ can further improve to $\approx 20000$. Consequently, the noise quantum number of the membrane thermal vibration and the shot noise in Eq.~(\ref{eq:SNR2}) are reduced by a factor of $\approx 10^{6}$.

The noise budget of the prospective EMO NMR detection is shown in Table~\ref{tb:p}, where all the aforementioned improvements are taken into account. The intrinsic Johnson noise aside, such improvements would lead to the the effective added noise temperature of the transducer of 6~K, outperforming the state-of-the-art low noise amplifier.


\section*{Definition of symbols}

\begin{table}[ht]
  \caption{Roman symbols (a -- c).} \label{tb:si1a}
  \begin{center}
   \begin{tabular}{|c|l|}
    \hline
   $a, a^{\dag}$ & LC annihilation and creation operators      \\
   $A$     & capacitor area      \\
   $b, b^{\dag}$ & membrane annihilation and creation operators    \\
   $B_{0}$               & magnetic field  \\
   $C$ & capacitance         \\
   $C_{\mathrm{em}}$   & electro-mechanical cooperativity        \\
   $C_{\mathrm{om}}$   & opto-mechanical cooperativity        \\
   \hline
 \end{tabular}
 \end{center}
\end{table}

\begin{table}[t]
  \caption{Roman symbols (d -- z).} \label{tb:si1b}
  \begin{center}
   \begin{tabular}{|c|l|}
   \hline
   $d$     & nominal capacitor gap      \\
   $d_{0}$ & equilibrium capacitor gap      \\
   $f_{\mathrm{in}}$ & mechanical thermal noise input      \\
   $g_{\mathrm{em}}$     &  single-photon electro-mechanical coupling rate     \\
   $g_{\mathrm{om}}$     &  single-photon opto-mechanical coupling rate     \\
   $G_{\mathrm{em}}$   & electro-mechanical coupling strength        \\
   $G_{\mathrm{om}}$   & opto-mechanical coupling strength        \\
   $l$    & nominal cavity length   \\
   $L$ & inductance              \\
   $m$ & membrane effective mass          \\
   $\mathcal{N}_{\mathrm{D}}$     &  intracavity photon number     \\
   $p$  &  linearized momentum     \\
   $P$ & membrane momentum         \\
   $P_{0}$  &  equilibrium momentum     \\
   $P_{\mathrm{D}}$ & LC drive power              \\
   $\mathcal{P}_{\mathrm{D}}$ & cavity drive power              \\
   $q$   &  linearized charge     \\
   $Q$ & charge        \\
   $Q_{0}$   & equilibrium charge    \\
   $q_{\mathrm{in}}$  & thermal charge fluctuation input     \\
   $Q_{\mathrm{in}}$  & charge fluctuation input      \\
   $S_{FF}$ & displacement noise spectral density              \\
   $S_{II}$ & current noise spectral density              \\
   $S_{\mathrm{oo}}$     & optical readout noise spectral density     \\
   $S_{qq}$     &  charge noise spectral density     \\
   $S_{VV}$ & voltage noise spectral density              \\
   $S_{XX}$ & optical shot noise spectral density             \\
   $S_{\dot{z}\dot{z}}$     & velocity noise spectral density      \\
   $T_{2}^{*}$ & NMR dephasing time constant              \\
   $T_{\mathrm{eff}}$     &  mechanical bath temperature     \\
   $T_{\mathrm{n}}$ & equivalent amplifier noise temperature              \\
   $V$ & voltage           \\
   $X_{\mathrm{in}}$   & optical cavity input      \\
   $X_{\mathrm{out}}$     &  optical cavity output     \\
   $z$  &  linearized displacement     \\
   $Z$ & membrane displacement         \\
   $Z_{0}$ &  equilibrium displacement     \\
   $z_{\mathrm{zpf}}$     & membrane zero-point fluctuation      \\

   \hline
 \end{tabular}
 \end{center}
\end{table}


\begin{table}[ht]
  \caption{Greek symbols.} \label{tb:si2}
  \begin{center}
   \begin{tabular}{|c|l|}
    \hline

   $\gamma_{\mathrm{i}}$ & LC dissipation rate              \\
   $\gamma_{\mathrm{m}}$ & mechanical dissipation rate              \\
   $\gamma_{\mathrm{o}}$ & optical dissipation rate              \\
   $\delta\omega$  & membrane frequency shift      \\
   $\Delta$ & tone offset frequency     \\
   $\eta_{\mathrm{p}}$ &  phase noise    \\
   $\kappa_{\mathrm{i}}$ & LC input coupling constant              \\
   $\kappa_{\mathrm{iT}}$ & net LC dissipation rate ($=\kappa_{\mathrm{i}}+\gamma_{\mathrm{i}}$) \\
   $\kappa_{\mathrm{o}}$ & optical output coupling constant              \\
   $\kappa_{\mathrm{oT}}$ & net optical dissipation rate ($=\kappa_{\mathrm{o}}+\gamma_{\mathrm{o}}$) \\
   $\phi$  &  linearized flux      \\
   $\Phi$ & flux        \\
   $\Phi_{0}$  & equilibrium flux      \\
   $\phi_{\mathrm{in}}$     &  thermal flux fluctuation input     \\
   $\Phi_{\mathrm{in}}$   &  flux fluctuation input     \\
   $\chi_{\mathrm{c}}$     &  cavity susceptibility     \\
   $\chi_{\mathrm{LC}}$     &  LC susceptibility     \\
   $\chi_{\mathrm{m}}$     &  mechanical susceptibility     \\
   $\Omega_{\mathrm{c}}$ & frequency of light        \\
   $\omega_{0}$ & unloaded membrane frequency        \\
   $\omega_{\mathrm{D}}$ & LC drive frequency       \\
   $\Omega_{\mathrm{D}}$ & cavity drive frequency       \\
   $\omega_{\mathrm{LC}}$ & resonance frequency of LC circuit        \\
   $\omega_{\mathrm{m}}$ & membrane resonance frequency    \\
   $\omega_{\mathrm{s}}$ & NMR frequency        \\

    \hline
  \end{tabular}
  \end{center}

\end{table}


\begin{thebibliography}{99}




  \bibitem{Polzik2014}
  T.~Bagci, A.~Simonsen, S.~Schmid, L.~G.~Villanueva, E.~Zeuthen, J.~Appel, J.~M.~Taylor, A.~S{\o}rensen, K.~Usami, A.~Schliesser, and E.~S.~Polzik, Optical detection of radio waves through a nanomechanical transducer, Nature \textbf{507}, 81--85 (2014).

  \bibitem{Bloch1946}
  F.~Bloch, W.~Hansen, and M.~Packard, The Nuclear Induction Experiment,
  Phys.~Rev.~\textbf{70}, 474--485 (1946).

  \bibitem{Purcell1946}
  E.~Purcell, H.~Torrey, and R.~Pound, Resonance Absorption by Nuclear Magnetic Moments in a Solid
  Phys.~Rev.~\textbf{69}, 37--38 (1946).

  \bibitem{Purcell1948}
  E.~Purcell, Nuclear Magnetism in Relation to Problems of the Liquid and Solid States, Science~\textbf{107}, 433--440 (1948).

  \bibitem{Bloch1946b}
  F.~Bloch, Nuclear Induction, Phys.~Rev.~\textbf{70}, 460--474 (1946).

  \bibitem{HB1997}
  D.~I.~Hoult and B.~Bhakar, NMR signal reception: Virtual photons and coherent spontaneous emission, Concepts~Magn.~Reson.~\textbf{9}, 277--297 (1997).

  \bibitem{Abragam}
  A.~Abragam, \textit{Principle of Nuclear Magnetism}, (Oxford University Press, 1961).

  \bibitem{Slichter2014}
  C.~P.~Slichter, The discovery and renaissance of dynamic nuclear polarization, Rep.~Prog.~Phys. \textbf{77}, 072501 (2014).

  \bibitem{Kikkawa2000}
  J.~M.~Kikkawa and D.~D.~Awschalom, All-optical magnetic resonance in semiconductors,
  Science \textbf{287}, 473--476 (2000).

  \bibitem{Savukov2006}
  I.~M.~Savukov, S.-K.~Lee, and M.~V.~Romalis, Optical detection of liquid-state NMR, Nature \textbf{442}, 1021--1024 (2006).

  \bibitem{PD2010}
  M.~Poggio and C.~L.~Degen, Force-detected nuclear magnetic resonance: Recent advances and future challenges, Nanotechnology \textbf{21}, 342001 (2010).

  \bibitem{Rugar2013}
  H.~J.~Mamin, M.~Kim, M.~H.~Sherwood, C.~T.~Rettner, K.~Ohno, D.~D.~Awschalom, and D.~Rugar, Nanoscale nuclear magnetic resonance with a nitrogen-vacancy spin sensor, Science \textbf{339}, 557--560 (2013).

  \bibitem{Wrachtrup2013}
  T.~Staudacher, F.~Shi, S.~Pezzagna, J.~Meijer, J.~Du, C.~A.~Meriles, F.~Reinhard, and J.~Wrachtrup, Nuclear magnetic resonance spectroscopy on a (5-nanometer)$^3$ sample volume, Science \textbf{339}, 561--563 (2013).

  \bibitem{Savukov2005}
  I.~M.~Savukov and M.~V.~Romalis, NMR detection with an atomic magnetometory, Phys.~Rev.~Lett. \textbf{94}, 123001 (2005).


  \bibitem{Hahn1950}
  E.~L.~Hahn, Spin Echoes, Phys. Rev. \textbf{80}, 580--594 (1950).

  \bibitem{Sillanpaa2011}
  F.~Massel, T.~T.~Heikkil\"a, J.~-M.~Pirkkalainen, S.~U.~Cho, H.~Saloniemi, P.~Hakonen, M.~A.~Sillanp\"a\"a, Microwave amplification with nanomechanical resonators, Nature \textbf{480}, 351--354 (2011).

  \bibitem{Taylor2011}
  J.~M.~Taylor, A.~S.~S{\o}rensen, C.~M.~Marcus, and E.~S.~Polzik, Laser Cooling and Optical Detection of Excitations in a LC Electrical Circuit, Phys.~Rev.~Lett. \textbf{107}, 273601 (2011).

  \bibitem{Wood2014}
  C.~J.~Wood, T.~W.~Borneman, and D.~G.~Cory, Cavity Cooling of an Ensemble Spin System, Phys.~Rev.~Lett. \textbf{112}, 050501 (2014). 

  \bibitem{Wood2016}
  C.~J.~Wood and D.~G.~Cory, Cavity Cooling to the Ground State of an Ensemble Quantum System, Phys.~Rev.~A \textbf{93}, 023414 (2016). 

  \bibitem{Clerk2010}
  A.~A.~Clerk, M.~H.~Devoret, S.~M.~Girvin, F.~Marquardt, and R.~J.~Schoelkopf, Introduction to quantum noise, measurement, and amplification, Rev.~Mod.~Phys. \textbf{82}, 1155--1208 (2010).

  \bibitem{Zeuthen2016}
  E.~Zeuthen, A.~Schliesser, A.~S.~S{\o}rensen, J.~M.~Taylor, Figures of merit for quantum transducers, arXiv:1610.01099.








  \bibitem{Qian2012}
  C.~Qian, J.~Murphy-Boesch, S.~Dodd, and A.~Koretsky, Sensitivity enhancement of remotely coupled NMR detectors using wirelessly powered parametric amplification, Mag. Reson. Med. \textbf{68}, 989--996 (2012). 

  \bibitem{Qian2013}
  C.~Qian, G.~Zabow, A.~Koretsky, Engineering novel detectors and sensors for MRI. J. Magn. Reson. \textbf{229}, 67--74 (2013). 

  \bibitem{Gigan2006}
  S.~Gigan, H.~R.~B\"{o}hm, M.~Paternostro, F.~Blaser, G.~Langer, J.~B.~Hertzberg, K.~C.~Schwab, D.~B\"{a}uerle, M.~Aspelmeyer, and A.~Zeilinger, Self-cooling of a micromirror by radiation pressure, Nature~\textbf{444}, 67--70 (2006). 

  \bibitem{Arcizet2006}
  O.~Arcizet, P.~F.~Cohadon, T.~Briant, M.~Pinard, and A.~Heidmann, Radiation pressure cooling and optomechanical instability of a micromirror, Nature~\textbf{444}, 71--74 (2006).

  \bibitem{Schliesser2006}
  A.~Schliesser, P.~Del'Haye, N.~Nooshi, K.~J.~Vahala, and T.~J.~Kippenberg, Radiation Pressure Cooling of a Micromechanical Oscillator Using Dynamical Backaction, Phys.~Rev.~Lett.~\textbf{97}, 243905 (2006).

  \bibitem{Griffin1993}
  L.~R.~Becerra, G.~J.~Gerfen, R.~J.~Temkin, D.~J.~Singel, and R.~G.~Griffin, Dynamic nuclear polarization with a cyclotron resonance maser at 5~T, Phys.~Rev.~Lett. \textbf{71}, 3561--3564 (1993).

  \bibitem{Schuster2010}
  D.~I.~Schuster, A.~P.~Sears, E.~Ginossar, L.~DiCarlo, L.~Frunzio, J.~J.~L.~Morton, H.~Wu, G.~A.~D.~Briggs, B.~B.~Buckley, D.~D.~Awschalom, and R.~J.~Schoelkopf, High-Cooperativity Coupling of Electron-Spin Ensembles to Superconducting Cavities, Phys.~Rev.~Lett. \textbf{105}, 140501 (2010).

  \bibitem{Kubo2010}
  Y.~Kubo, F.~R.~Ong, P.~Bertet, D.~Vion, V.~Jacques, D.~Zheng, A.~Dr{\'e}au, J.~-F.~Roch, A.~Auffeves, F.~Jelezko, J.~Wrachtrup, M.~F.~Barthe, P.~Bergonzo, and D.~Esteve, Strong Coupling of a Spin Ensemble to a Superconducting Resonator, Phys.~Rev.~Lett.~\textbf{105}, 140502 (2010).

  \bibitem{Abe2011}
  E.~Abe, H.~Wu, A.~Ardavan, and J.~J.~L.~Morton, Electron spin ensemble strongly coupled to a three-dimensional microwave cavity, Appl.~Phys.Lett. \textbf{98}, 251108 (2011).

  \bibitem{Bienfait2016}
  A.~Bienfait, J.~J.~Pla, Y.~Kubo, X.~Zhou, M.~Stern, C.~C.~Lo, C.~D.~Weis, T.~Schenkel, D.~Vion, D.~Esteve, J.~J.~L.~Morton, and P.~Bertet, Controlling spin relaxation with a cavity, Nature \textbf{531}, 74--77 (2016).

  \bibitem{Eichler2017}
  C.~Eichler, A.~J.~Sigillito, S.~A.~Lyon, and J.~R.~Petta, Electron Spin Resonance at the Level of 10$^4$ Spins Using Low Impedance Superconducting Resonators, Phys.~Rev.~Lett.~\textbf{118}, 037701 (2017).


  \bibitem{Butler2011}
  M.~C.~Butler and D.~P.~Weitekamp, Polarization of nuclear spins by a cold nanoscale resonator, Phys.~Rev.~A \textbf{84}, 063407 (2011). 

  \bibitem{Mehring}
  M.~Mehring, \textit{Principles of High-Resolution NMR in Solids, Second Edition}, (Springer-Verlag, 1983).

  \bibitem{Takeda2012}
  K.~Takeda, N.~Ichijo, Y.~Noda, and K.~Takegoshi, Elemental analysis by NMR, J. Magn. Reson. \textbf{224}, 48--52 (2012).

  \bibitem{Yamada2015}
  K.~Yamada, K.~Kitagawa, and M.~Takahashi, Field-swept $^{33}$S NMR study of elemental sulfur, Chem. Phys. Lett. \textbf{618}, 20--23 (2015). 


  \bibitem{Takeda2008}
  K.~Takeda, OPENCORE NMR: open-source core modules for implementing an integrated FPGA-based NMR spectrometer, J. Magn. Reson. \textbf{192}, 218--229 (2008).



  \bibitem{Hisatomi2016}
  R.~Hisatomi, A.~Osada, Y.~Tabuchi, T.~Ishikawa, A.~Noguchi, R.~Yamazaki, K.~Usami, and Y.~Nakamura, Bidirectional conversion between microwave and light via ferromagnetic magnons, Phys.~Rev.~B \textbf{93}, 174427 (2016).

  \bibitem{Karrai2004}
  C.~Hohberger~Metzger and K.~Karrai, Cavity cooling of a microlever, Nature~\textbf{432}, 1002--1005 (2004).


\end{thebibliography}
\end{document}